\def\Nc{N_{\rm c}}
\def\q{{\bm q}}
\def\x{{\bm x}}

\def\Re{\operatorname{Re}}
\def\Im{\operatorname{Im}}

\def\zero{\mathsf{0}}
\def\one{\mathsf{1}}
\def\two{\mathsf{2}}
\def\three{\mathsf{3}}
\def\five{{\mathsf{5}}}

\def\zh{z_{\rm h}}
\def\Rads{\mathtt{R}}

\def\alphas{\alpha_{\rm s}}
\def\lstop{\ell_{\rm stop}}
\def\lmax{\ell_{\rm max}}
\def\tension{{\mathbb T}}
\def\lgrav{\Sigma}
\def\dof{x}

\documentclass[prd,preprint,eqsecnum,nofootinbib,amsmath,amssymb,
               tightenlines,dvips]{revtex4}
\usepackage{graphicx}
\usepackage{bm}
\usepackage{amsfonts}
\usepackage{amssymb}

\begin {document}



\title
    {
      Tidal stretching of gravitons into classical strings:
      application to jet quenching with AdS/CFT
    }

\author{
  Peter Arnold, Phillip Szepietowski, Diana Vaman, and Gabriel Wong
}
\affiliation
    {%
    Department of Physics,
    University of Virginia, Box 400714,
    Charlottesville, Virginia 22904, USA
    }%

\date {\today}

\begin {abstract}%
{%
  Previous work has shown that the standard
  supergravity approximation can break
  down when using AdS/CFT duality to study certain top-down
  formulations of the
  jet stopping problem in strongly-coupled ${\cal N}{=}4$
  super-Yang-Mills (SYM) plasmas, depending on the virtuality of the
  source of the ``jet.''  In this paper, we identify the nature
  of this breakdown:
  High-momentum gravitons in the gravitational dual
  get stretched into relatively
  large classical string loops by tidal forces associated with the
  black brane.  These stringy excitations of the graviton are not
  contained in the supergravity approximation, but we show that
  the jet stopping problem can nonetheless still be solved by
  drawing on various string-theory methods
  (the eikonal approximation, the Penrose limit, string quantization in
  pp-wave backgrounds) to obtain
  a probability distribution for the late-time classical string
  loops.
  In extreme cases, we find that the gravitons are stretched into
  very long folded strings which are qualitatively similar to the
  folded classical strings originally used by Gubser, Gulotta,
  Pufu and Rocha to model the jet stopping problem.
  This makes a connection in certain cases
  between the different methods that have
  been used to study jet stopping with AdS/CFT
  and gives a specific example
  of a precise ${\cal N}{=}4$ SYM problem that generates such strings
  in the gravity description.
}%
\end {abstract}

\maketitle
\thispagestyle {empty}


\section {Introduction}
\label{sec:intro}

\subsection {Background}

Inspired by the observation of (and rapidly growing body of experimental
information on) jet quenching in relativistic heavy ion collisions,
there has for many years been an interest in the theory of jet quenching
and what can be learned about that theory by studying interesting
limiting cases.  One of the simplest-to-pose thought experiments is
this: How far does a very-high momentum excitation (the potential
precursor of a would-be jet) travel in a thermal QCD medium before
it loses energy, stops, and thermalizes in the medium?  And how does
the answer to that question depend on the effective
strength $\alphas$ of the strong coupling?

This question can be addressed from first principles in various
theoretical limits.  One such limit is that of weak coupling,
which in principle applies to asymptotically large temperatures $T$ and
jet energies $E$, for which the relevant running values of $\alphas$
are small.  In that limit, the stopping distance
$\lstop$ for a high-energy
parton ($E \gg T$) scales with energy as $E^{1/2}$, up to logarithms.%
\footnote{
  A specific weak-coupling calculation
  of the stopping distance for QCD in
  the high-energy limit may be found in ref.\ \cite{stop}.
  However, the scaling of this result was
  implicit in the early pioneering work of
  refs.\ \cite{BDMPS,Zakharov} on
  bremsstrahlung and energy loss rates in QCD plasmas.
  Introducing supersymmetry will not change the
  conclusion that the stopping distance scales as $E^{1/2}$ (up to
  logarithms) at weak coupling.
}
A contrasting limit of interest occurs when the running values of
$\alphas$ relevant to jet stopping are all large.%
\footnote{
  In a weak-coupling analysis, the two running couplings relevant to
  jet stopping are, roughly,
  $\alphas(T)$ and $\alphas(Q_\perp)$, where
  $Q_\perp \sim (\hat q E)^{1/4}$ grows slowly with energy and is the
  scale of the typical relative momentum between two daughter
  partons when a high energy parton splits through hard bremsstrahlung
  or pair production.
  ($\hat q \sim \alphas^2 T^3$ is a scale characteristic of the plasma
  that parametrizes transverse momentum diffusion of high-energy partons.)
  A third limiting case of interest, not addressed
  in this paper, is where $\alphas(T)$ is large but $\alphas(Q_\perp)$
  is small.  See, for example, Liu, Rajagopal, and Wiedemann \cite{LRW}.
}
This problem is
not very tractable from first principles in QCD itself,
but, through gauge-gravity duality,
progress can be made for QCD-like plasmas with gravity duals, such as
${\cal N}{=}4$ super Yang Mills (SYM) theory.
For some years, people have considered various ways to study
analogs of jet stopping in such plasmas, namely the stopping distance
for various types of localized, high-momentum excitations.
The exact stopping distance depends on details of exactly how the
``jet'' is prepared, but universally these studies have found that
the maximum possible stopping distance $\lmax$ scales with energy as
$E^{1/3}$ \cite{GubserGluon,HIM,CheslerQuark,adsjet,adsjet2,CHR},
in contrast to the weak-coupling scaling of $E^{1/2}$.
This is an interesting theoretical result because it teaches us
that the scaling of jet stopping with energy depends on the strength
of the coupling.  It remains an open question (which we will not
answer here) how $E^{1/3}$ starts to
move toward $E^{1/2}$ as one lowers the coupling, and vice versa.%
\footnote{
  See the conclusion of ref.\ \cite{R4} for further discussion of
  this point.
}

The stopping distance of high-momentum, localized excitations
traveling through the plasma depends on more than just the energy of
the excitation.  Depending on exactly how one creates the excitation
(the ``jet''), one may get stopping distances $\lstop$ significantly
smaller than the maximum $\lmax$.  As an example from weak coupling,
imagine that we spread out the total energy and momentum $E$ of the
jet among 10 partons, each having energy $E/10$, rather than putting
it all into a single
parton of energy $E$.  Each of the 10 partons has lower energy than
the single one and so will stop sooner; so the stopping distance for
the high-momentum excitation depends on how many high-energy partons
we use in the initial state.  In the weak-coupling case, the maximum
stopping distance $\ell_{\rm max}$ corresponds to the particular
initial state where all the energy is concentrated into a single
initial parton.

\begin {figure}
\begin {center}
  \includegraphics[scale=0.35]{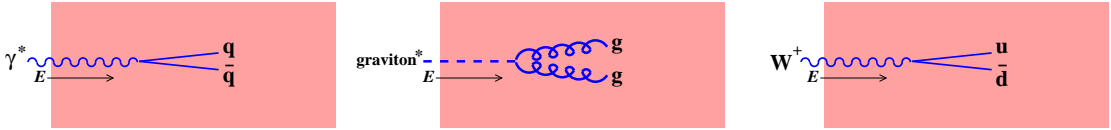}
  \caption{
     \label{fig:Wdecay}
     Examples of the decay of a very high-energy
     (a) slightly virtual photon, (b) slightly virtual graviton, or
     (c) on-shell $W^+$ boson, inside a standard-model
     quark-gluon plasma, producing high-momentum partons moving
     to the right.
     In the context of ${\cal N}{=}4$ super Yang Mills,
     the $q$, $u$, and $\bar d$ above
     represent adjoint-color fermions or scalars carrying R charge.
     For strong coupling, of course, one should not picture perturbatively,
     as in this figure, the
     high-momentum excitation created in the plasma by the decay.
  }
\end {center}
\end {figure}

In the strong-coupling case, we cannot speak of individual partons,
but the stopping distance again depends on how we prepare the initial
high-momentum excitation.  In our work \cite{adsjet,adsjet2,R4}, we
create the initial excitation in a way that is analogous to what you
would get if a high-momentum, slightly virtual photon (or graviton or
other massless particle)
decayed hadronically in the quark-gluon plasma, as depicted in
fig.\ \ref{fig:Wdecay}.
Alternatively, one could consider the decay of a high-momentum on-shell W
boson (also depicted).
For these methods of creating ``jets,'' one finds
that the maximum possible stopping distance scales as
\begin {equation}
   \lmax \sim \frac{E^{1/3}}{T^{4/3}} \,.
\label {eq:lmax}
\end {equation}
As we will review later, it turns out that
the stopping distance may be made smaller than (\ref{eq:lmax}) by varying
the virtuality
$-q^2 \equiv -q_\mu q^\mu$ of the virtual photon (or equivalently the
mass-squared $M_{\rm w}^2$ of the on-shell W boson)
\cite{HIM,adsjet2}.
The important point is that there is a range of stopping
distances $\lstop \lesssim \lmax$ for our ``jets,'' depending on the details
of how those excitations are created.

\begin {figure}
\begin {center}
  \includegraphics[scale=0.35]{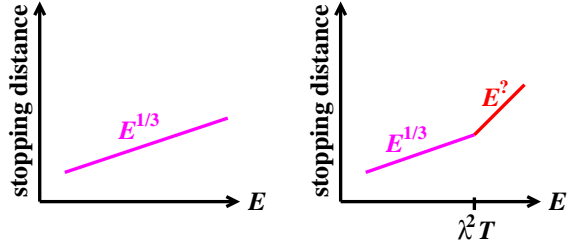}
  \caption{
     \label{fig:stopvE}
     Examples of two different scenarios for the high-energy ($E \gg T$)
     behavior of
     the maximum jet stopping distance $\lmax(E)$ which are indistinguishable
     with $\lambda{=}\infty$ calculations.
  }
\end {center}
\end {figure}

\begin {figure}
\begin {center}
  \includegraphics[scale=0.8]{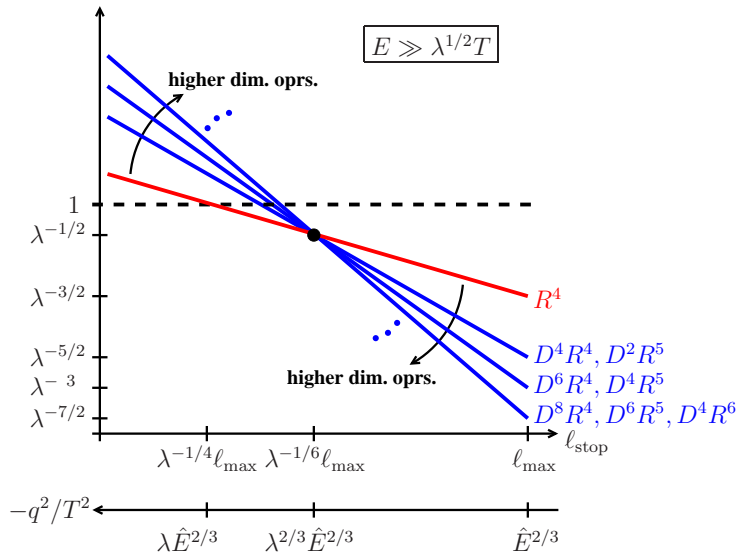}
  \caption{
     \label{fig:R4}
     A parametric picture of the relative importance of
     higher-derivative corrections to the low-energy
     supergravity action as a function of the stopping distance
     $\ell_{\rm stop}$
     (using the $\lambda{=}\infty$ result for $\ell_{\rm stop}$).
     The axis are both logarithmic, and an importance of $1$ indicates
     that the individual correction would, by itself, significantly modify
     the $\lambda{=}\infty$ analysis.  The measure of ``importance'' is
     explained in ref.\ \cite{R4}.  Also shown, as an alternative
     horizontal axis, is the 4-dimensional virtuality $-q^2$ of
     the source that created the jet, where ${\hat E}\equiv E/T$.
  }
\end {center}
\end {figure}

Most top-down studies of jet stopping using gauge-gravity duality
have studied the infinite color and infinite coupling limit,
$N_{\rm c}{=}\infty$ and $\lambda{=}\infty$, where
$\lambda \equiv N_{\rm c} g_{\rm YM}^2$ is the 't Hooft coupling.
To understand the true high-energy behavior, however, it is important
to study the corrections to these limits.
As an example, fig.\ \ref{fig:stopvE} shows two different
scenarios one might imagine for the maximum stopping distance
$\lmax$ for strongly-coupled ${\cal N}{=}4$ SYM.
One is that $\lmax$ grows like $E^{1/3}$ at high energy, up to
arbitrarily high energies.  The other is that it starts growing like
$E^{1/3}$ at high energy $E \gg T$, but then crosses over to some
different power-law behavior once $E$ exceeds some positive power
of $\lambda$ times $T$ (e.g.\ $\lambda^2 T$ in the figure).
In the latter case, $E^{1/3}$ would not be the true behavior for arbitrarily
large $E$ and large but finite $\lambda$.  But there is no way to
tell the difference between these two scenarios if one only
has $\lambda{=}\infty$ calculations!  For this reason, three of us
analyzed the parametric size of finite-$\lambda$ corrections to jet
stopping distances in ref.\ \cite{R4}.  We found that the formal
expansion in $1/\sqrt\lambda$ (which corresponds
to an expansion in the string parameter $\alpha'$ on the gravity side)
breaks down for some jets and is safe for
others, depending on the stopping distance $\lstop$ of the jet
(and therefore on the virtuality $-q^2$), as depicted in
fig.\ \ref{fig:R4}.
The stopping distance and virtuality parametrize the horizontal
axis in this figure.  The vertical axis is a relative measure of the
importance of a given correction compared to the $\lambda{=}\infty$
result (see ref.\ \cite{R4} for details).
The curves are labeled by the sequence of higher-curvature
terms in the gravitational dual theory action that correspond,
via the AdS/CFT correspondence, to a sequence of corrections in powers
of $1/\sqrt\lambda$ in the 3+1 dimensional ${\cal N}{=}4$ SYM theory.
Throughout, $N_{\rm c}$ is taken to be infinite.  The result of this
study was that, for $\lambda \gg 1$, corrections to the
$\lambda{=}\infty$ result are parametrically
small for $\lambda^{-1/6}\lmax \ll \lstop \lesssim \lmax$.
In particular, corrections to the maximum stopping distance
$\lmax \propto E^{1/3}$ are small.  But the interesting case
is when jets are created in such a way that
\begin {subequations}
\label {eq:fun}
\begin {equation}
   T^{-1} \ll \lstop \lesssim \lambda^{-1/6} \lmax ,
\label {eq:lstopwindow}
\end {equation}
which is
\begin {equation}
   T^{-1} \ll 
   \lstop \lesssim \frac{\bigl(E/\sqrt\lambda\>\bigr)^{1/3}}{T^{4/3}} \,.
\end {equation}
\end {subequations}
In this case, the fate of $\lambda{=}\infty$ results for
the stopping distance was unclear.
For $\lstop \sim \lambda^{-1/6} \lmax$, all the corrections are the same
size, and so the formal expansion in powers of $1/\sqrt\lambda$ has broken
down.  Yet the individual corrections are all small (of relative importance
$\lambda^{-1/2}$) for that $\lstop$.  From fig.\ \ref{fig:R4}, we cannot
tell whether the {\it sum}\/ of the corrections to $\lambda{=}\infty$
will remain small for $\lstop \lesssim \lambda^{-1/6}\lmax$ or whether,
instead, the $\lambda{=}\infty$ calculation becomes useless there.

The purpose of the present paper is to understand the physics (on the
gravity side) of what is going on in the region (\ref{eq:fun}) where
the naive expansion in powers of $1/\sqrt\lambda$ (powers of $\alpha'$)
breaks down, and to figure out how to account for the effect of this physics
on the jet stopping distance.
Note that the interesting window (\ref{eq:lstopwindow}) of stopping
distances exists only if
the energy is large enough that $T^{-1} \ll \lambda^{-1/6} \lmax$.
By (\ref{eq:lmax}), this requires $E \gg \lambda^{1/2} T$, which we
will assume throughout the rest of this paper.

\begin {figure}
\begin {center}
  \includegraphics[scale=0.5]{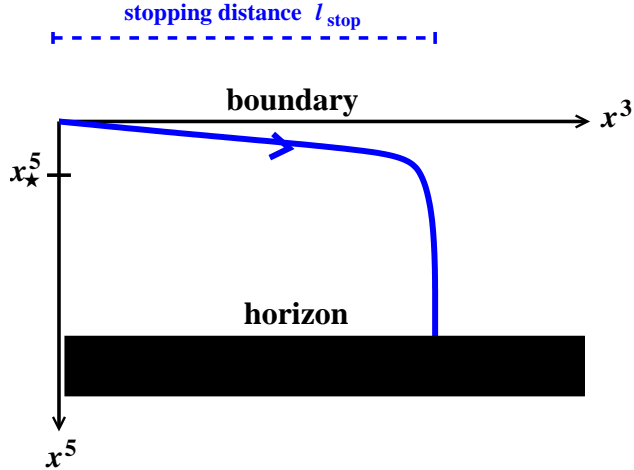}
  \caption{
     \label{fig:fall}
     Qualitative sketch of the motion though AdS$_5$-Schwarzschild
     of a wave packet with high 3-momentum in the $x^\three$ direction.
     As measured by $x^\zero$, the particle
     takes infinitely long to reach the horizon.
     Of special importance is the parametric
     scale $x^\five_\star$ in the fifth
     dimension, where the trajectory turns over and beyond which
     progress in
     $x^\three$ rapidly slows to a stop.
  }
\end {center}
\end {figure}

Before outlining what we have done, it will be useful to first explain
one other qualitative feature of the $\lambda{=}\infty$ calculation.
Excitations created in the field theory correspond
to excitations created on the boundary of AdS$_5$-Schwarzschild,
which then fall towards the black brane over time, such as
depicted in fig.\ \ref{fig:fall}.  The 3-space distance
that this excitation travels
before falling into the horizon matches the
stopping distance of the corresponding excitation in
${\cal N}{=}4$ SYM.%
\footnote{
  See ref.\ \cite{adsjet2} for a discussion in the context of the
  present paper, but this correspondence is implicit in the earlier
  work of refs.\ \cite{GubserGluon,HIM,CheslerQuark}.
}
For $\lstop \ll \lmax$, which includes the region (\ref{eq:fun})
of interest, there is a nice simplification.
On the gravity side, the excitation falling in fig.\ \ref{fig:fall}
turns out to be a spatially {\it small}\/
wavepacket which can be treated in the geometric optics approximation.
The wavepacket's motion is the same (up to parametrically small
corrections) as that of a 5-dimensional ``particle'' traveling in
the AdS$_5$-Schwarzschild geometry, and so it follows a geodesic whose
trajectory is easily calculated in terms of the 4-momentum $q_\mu$ of
the excitation.  (See section \ref{sec:geodesic} for more detail.)

It will be important for what follows to remember that
the AdS/CFT correspondence is really
a correspondence between field theory and 10-dimensional string theory.
In the strong-coupling limit $\lambda{=}\infty$ of the field theory,
the general correspondence reduces to one between the field theory and the
infrared limit of the string theory, which is a supergravity
theory.  The quanta of the supergravity fields
correspond to string states that are massless in flat 10-dimensional
spacetime, such as the graviton.
For $\lambda{=}\infty$, the well-known
gravitational dual of finite-temperature ${\cal N}=4$ SYM is
Type IIB supergravity in an (AdS$_5$-Schwarzschild)$\times S^5$
background.


\subsection {What we find}

The classical wave packet falling in fig.\ \ref{fig:fall} is a localized,
classical excitation of the supergravity fields.
For the sake of
specificity, consider the case where it is
an excitation of the background gravitational field.
In general, a classical wave is a coherent superposition of the
corresponding quanta, and so our high-frequency
classical wavepacket is a coherent
superposition of high-frequency gravitons.
Since our wavepacket behaves like a particle (in the
geometric optics approximation appropriate for $\lstop \ll \lmax$),
let's follow just one of these gravitons.  So think of the trajectory in
fig.\ \ref{fig:fall} as
that of a single high-momentum graviton with a localized wave function.
As we will discuss later, interactions between the gravitons that make
up the wavepacket are very small, and so it is adequate to think
about the evolution of individual high-momentum gravitons.

A graviton is really a tiny loop of string whose internal degrees of freedom
are in their ground state.
Because of the gravitational field from the black brane, this closed
string will feel tidal forces as it falls, which will try to stretch
the string in some directions and squeeze it in others.
As the graviton gets further from the boundary (and so closer to
the black brane), the tidal forces will increase, and eventually
they will become large enough to excite the internal string degrees
of freedom of the graviton.
We will argue that it is the excitation
of these string degrees of freedom that is responsible for the
breakdown of the expansion in fig.\ \ref{fig:R4} in the problem
region (\ref{eq:fun}).

We will find that, in the problematic case (\ref{eq:fun})
where $\lstop \ll \lambda^{-1/6}\lmax$,
the tidal forces are strong enough to stretch that loop
of string to become classically large before the stopping distance
is reached.  This is why stringy corrections cannot be ignored
in that case, explaining the breakdown of the expansion in
fig.\ \ref{fig:R4}.
(In contrast, the tidal forces are
not strong enough to excite the graviton's internal degrees of freedom
soon enough when $\lstop \gg \lambda^{-1/6}\lmax$.)
Though the resulting classical string loop will be large
compared to the size of a graviton, we should ask how its
size compares to the stopping distance $\lstop$.  We will find
that the ratio of (i) the stretched, classical string's size
in the direction of motion $x^\three$ to
(ii) the stopping distance $\lstop$ is parametrically
of order
\begin {equation}
   \frac{(\delta x^\three)_{\rm string}}{\lstop}
   \sim
   \lambda^{-1/4} \, \ln^{1/2}\biggl(\frac{\lambda^{-1/6} \lmax}{\lstop}\biggr)
   .
\label {eq:ratio}
\end {equation}
Because of the $\lambda^{-1/4}$, this ratio
is typically parametrically small
for large but finite $\lambda$, and we will argue that the stretching
of the graviton into a string (and the accompanying breakdown of
the formal expansion in $1/\sqrt{\lambda}$ in fig.\ \ref{fig:R4})
then has sub-leading impact on
$\lambda{=}\infty$ results for the stopping distance.
But (\ref{eq:ratio}) also includes cases where the stretching of the
string may play an important role:
If one considers a situation where the argument of the logarithm in
(\ref{eq:ratio}) is exponentially large,
then the logarithm can be large enough to compensate for
the factor of $\lambda^{-1/4}$.  We discuss this situation further
in our conclusions, where we make contact with folded classical
string configurations that were originally used by
Gubser et al.\ \cite{GubserGluon} to model jet stopping.

\begin {figure}
\begin {center}
  \includegraphics[scale=0.5]{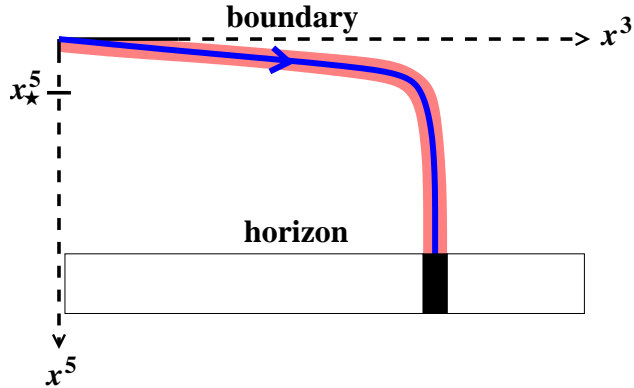}
  \caption{
     \label{fig:fallpp}
     The (pink) shaded area represents
     a narrow region of space-time around the null geodesic of
     fig.\ \ref{fig:fall}.  The AdS$_5$-Schwarzschild metric in this
     region may be approximated as a
     pp-wave background for the purpose of quantizing a small,
     falling loop of string that describes a graviton (or other
     particle) in the initial falling wavepacket.
  }
\end {center}
\end {figure}

Since the tidal forces stretch a quantum string (the graviton) into
a larger classical string, one may wonder whether or not it is possible
to do a real, detailed calculation of the transition between the two.
Having restricted attention
to a single graviton, our problem reduces to following the evolution
of a single closed string in the AdS$_5$-Schwarzschild background.  In
general, it is not known (for all practical purposes)
how to quantize a string in an
AdS$_5$-Schwarzschild background.  But remember that our graviton is
localized and so only probes a region of space-time near the geodesic
depicted in fig.\ \ref{fig:fall}.  It is enough to consider only a
narrow region of the space-time that lies near a null geodesic, as
depicted in fig.\ \ref{fig:fallpp}, and so we may treat the full
background metric in an approximation (known as a Penrose limit) that
treats displacements from the null geodesic as small.  The resulting
approximation to the background metric is an example of what is known
as a pp-wave background, and it {\it is}\/ known how to quantize a
string in a pp-wave background.  In particular, it will be possible
to calculate the probability distribution of the shape of the
classical string loop.  The methods we use are similar to
previous works by other authors on the excitation of string modes
in scattering processes and/or in pp-wave backgrounds
\cite{Veneziano,PRT,HS,GPZS,dVS}.

\medskip

In the next section, we give a simple, back-of-the-envelope argument
that yields the result $(\delta x^\three)/\ell_{\rm stop} \sim \lambda^{-1/4}$
but is not precise enough to explain the logarithmic factor in
(\ref{eq:ratio}).  Back-of-the-envelope estimates are sometimes
enlightening and sometimes frustratingly unconvincing, and
the rest of the paper is devoted to a more formal calculation
along the lines just described.  We set up our notational conventions
and review $\lambda{=}\infty$ calculations in section \ref{sec:setup}.
Next we review in section \ref{sec:R4review}
the source of the breakdown of the $\alpha'$ expansion
for $\lstop \ll \lambda^{-1/6} \lmax$, depicted in fig.\ \ref{fig:R4},
and use it to motivate why these problems should be overcome by
following the quantum evolution of single closed strings in the
AdS$_5$-Schwarzschild background.
In section \ref{sec:Penrose}, we take the relevant Penrose limit
of the AdS$_5$-Schwarzschild metric, crudely depicted in
fig.\ \ref{fig:fallpp}.  We are then ready to quantize the
string in section\ \ref{sec:quantize} and solve for the evolution
of the graviton into a classical loop of string.  We then compute
the average size of the resulting classical loops of string, which
parametrically gives (\ref{eq:ratio}), including the logarithmic factor,
for the interesting case $\lstop \ll \lambda^{-1/6} \lmax$.
Some readers may wonder how a quantum treatment of
the stretching of gravitons can underlie the analysis, given that the
gravitational/string dual theory is supposed to be classical in
the limit $\Nc{=}\infty$ taken in this paper.  We discuss this
in section \ref{sec:gravitons}, where we also give a more
detailed justification for treating the gravitons in the wavepacket
as independent.
In section \ref{sec:PenroseCheck}
(supplemented by Appendix \ref{app:PenroseCheck}),
we then revisit the Penrose limit
used in our analysis and verify that it is justified, provided
that (\ref{eq:ratio}) is small.
Finally, we offer our conclusions in section \ref{sec:conclusion}.


\section{A back-of-the-envelope estimate}
\label {sec:envelope}

In this section, we will make a parametric estimate of the amount of
tidal stretching of the string compared to the size of the stopping
distance $\lstop$.  In a later section, we will review the
$\lambda{=}\infty$ results for how the stopping distance depends on
the energy and 4-virtuality of our jet source, but here the only
thing we will need to know is that the stopping distance given by
following a null geodesic as in fig.\ \ref{fig:fall} is proportional
to a power of the slope $dx^\three/dx^\five$ of that geodesic where it starts,
at the boundary.  The more downward-directed one starts the trajectory
in fig.\ \ref{fig:fall}, the less distance it will travel in $x^\three$
before reaching the horizon.

Now interpret the trajectory of fig.\ \ref{fig:fall} as a trajectory
for the center of mass of a tiny, falling loop of string.
Once the string gets far enough from the
boundary that the tidal forces dominate over the string tension,
then the string tension becomes ignorable, and different pieces of
the string will fall independently along their own geodesics,
the string stretching accordingly.
Imagine plotting two such geodesics, for the two bits of the
string loop that are most separated.  The separation of
those geodesics is a measure of the extent of the tidally-stretched
loop of string as it falls towards the horizon.
The {\it proper} size of the string should start out of order the
quantum mechanical size $\lgrav$ of the graviton, which is
roughly set by dimensional
analysis in terms of the string tension $\tension$ as
\begin {equation}
   \lgrav_{\rm graviton} \sim \tension^{-1/2} \sim \sqrt{\alpha'} ,
\label {eq:lgrav}
\end {equation}
where $\alpha' = 1/2\pi\tension$ is the string slope parameter.

\begin {figure}
\begin {center}
  \includegraphics[scale=0.3]{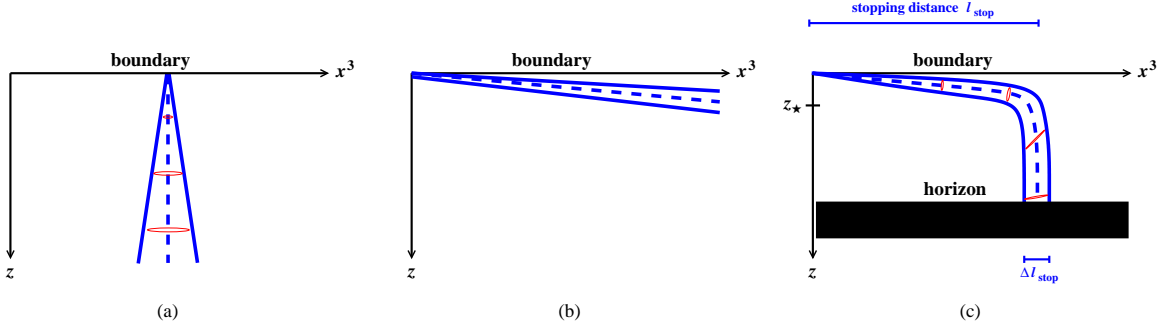}
  \caption{
     \label{fig:boost}
     (a) Parallel geodesics in AdS$_5$ in the Lorentz frame where the
     excitation is at rest in 3-space.  These geodesics maintain a constant
     proper separation as they fall into the bulk, and this
     separation should be thought of as of order the characteristic
     size ($\sim \sqrt{\alpha'}$)
     of the closed quantum string loop describing the graviton (or
     other massless string mode).  The narrow red loops are meant to
     be suggestive of the closed string loop.
     (b) The same picture boosted to the original Lorentz frame.
     (c) A picture of how those geodesics evolve in AdS$_5$-Schwarzschild
     rather than AdS$_5$.  The early-time behavior is the same as (b).
     [For classical oscillating string solutions, the strings depicted
     in (a) may be thought of as snapshots at moments when the string's
     proper extent in $x^\three$ is at, say,
     maximum (or half-maximum or whatever).
     Such solutions would similarly oscillate in the
     $z \ll z_\star$ part of (b) but not in the $z \gg z_\star$ part,
     where tidal forces dominate over tension.]
  }
\end {center}
\end {figure}

Very close to the boundary, the tidal forces due to the black hole are
negligible, and the closed loop of string is in its ground state.
We can set up our two geodesics above so that, correspondingly, they
maintain constant proper separation $\lgrav_{\rm graviton}$
near the boundary, where
the AdS$_5$-Schwarzschild metric approaches a purely AdS$_5$ metric.
To see how this works, imagine making a 4-dimensional boost from (i)
the plasma rest frame, in which we create an excitation with
large 4-momentum
$q^\mu = (\omega,0,0,q_\three) \simeq (E,0,0,E)$ and relatively
small 4-virtuality $-q^2 \ll E^2$, to (ii) the
excitation's initial rest frame, where the 4-momentum is
instead $(\sqrt{-q^2},0,0,0)$.  The Lorentz boost factor for this
transformation is
\begin {equation}
   \gamma = \sqrt{\frac{\omega^2}{-q^2}}
   \simeq \sqrt{\frac{E^2}{-q^2}}
   \gg 1 .
\label {eq:gamma}
\end {equation}
In AdS$_5$, the trajectory in the new frame will drop straight down away
from the boundary,
as depicted by the dashed line in fig.\ \ref{fig:boost}a.

Now consider the graviton as an extended object with proper
size $\lgrav$.  The two straight solid null lines in fig.\ \ref{fig:boost}a
depict the extent of the graviton in AdS$_5$ in the excitation's rest
frame at early times.
In Poincare coordinates for pure AdS$_5$,
\begin {equation}
   (ds)^2 =
   \frac{\Rads^2}{z^2} \,\bigl (dx^\mu \eta_{\mu\nu} dx^\nu + (dz)^2 \bigr) ,
\end {equation}
null geodesics are straight lines.  Here $\Rads$ is the radius of
the 5-sphere $S^5$.
We parametrize the two solid lines of fig.\ \ref{fig:boost}a as
\begin {equation}
   x^I = \bigl(\gamma_+,0,0,\pm\beta_+\gamma_+,1\bigr) \, z
\label {eq:xline}
\end {equation}
with $\beta_+ \ll 1$ and $\gamma_+ \equiv (1-\beta_+^2)^{-1/2} \simeq 1$.
Because of the warp factor in the metric, these two lines are parallel
and maintain constant proper separation
\begin {equation}
  \sqrt{\Delta x^\three \, g_{\three\three} \, \Delta x^\three}
  = 2\beta_+ \gamma_+ \Rads
  \simeq 2\beta_+ \Rads
\label {eq:separation}
\end {equation}
as a function of the rest-frame time.
Setting this proper separation
to be of order the graviton size $\lgrav$
given by (\ref{eq:lgrav}) then gives
\begin {equation}
   \beta_+ \sim
   \frac{\lgrav_{\rm graviton}}{\Rads} \sim \frac{\sqrt{\alpha'}}{\Rads} \,.
\end {equation}
The AdS/CFT dictionary relates $\alpha'$ in the string theory with
the 't Hooft coupling of
${\cal N}{=}4$ SYM by \cite{MAGOO}
\begin {equation}
   \frac{\alpha'}{\Rads^2} = \lambda^{-1/2}
   \equiv (g_{\rm YM}^2 N_{\rm c})^{-1/2} .
\label {eq:lambda}
\end {equation}
So
\begin {equation}
   \beta_+ \sim \lambda^{-1/4} ,
\end {equation}
and (\ref{eq:xline}) gives
\begin {equation}
   x^I \sim \bigl(1,0,0,\pm\lambda^{-1/4},1\bigr) \, z .
\end {equation}

Now boost back to the original plasma frame using
(\ref{eq:gamma}) to get the early-time trajectories depicted by solid lines
in fig.\ \ref{fig:boost}b:
\begin {equation}
   x^I \sim
   \Bigl(\gamma(1 \pm \lambda^{-1/4}),0,0,\gamma(1\pm\lambda^{-1/4}),1\Bigr) \, z
   ,
\end {equation}
where we have used $\gamma \gg 1$ (\ref{eq:gamma}).
Then
\begin {equation}
  \frac{dx^\three}{dz} \Bigg|_{\rm initial} \simeq \gamma(1\pm\lambda^{-1/4}) ,
\end {equation}
which represents a small relative variation
\begin {equation}
  \frac{\Delta(dx^\three/dz)}{dx^\three/dz} \Bigg|_{\rm initial}
  \sim \lambda^{-1/4}
\label {eq:relative}
\end {equation}
in the initial slope $dx^\three/dz$ of the trajectory.
As discussed before, the stopping distance (which requires
a calculation in the full AdS$_5$-Schwarzschild metric) covered
by a null geodesic is power-law related to this initial slope,
and so the difference $\Delta \lstop$ in how far the two
bits of string travel also has the same small size (\ref{eq:relative})
relative to $\lstop$:
\begin {equation}
  \frac{\Delta\lstop}{\lstop} \sim \lambda^{-1/4} .
\end {equation}
This is just our parametric result (\ref{eq:ratio}) quoted in the
introduction but without the logarithmic factor.
The logarithmic factor requires a more detailed analysis.


\section {Setup}
\label {sec:setup}

\subsection {Notation}

In this paper, we will use Greek letters for 4-dimensional space-time indices
($\mu,\nu={\mathsf{0,1,2,3}}$) and upper-case roman letters
for indices that run over all 5 dimensions of AdS$_5$-Schwarzschild
($I,J={\mathsf{0,1,2,3,5}}$).
One form of the metric we use for AdS$_5$-Schwarzschild is
\begin {equation}
   (ds)^2 = \frac{\Rads^2}{z^2} \bigl[-f (dt)^2 + (d\x)^2 + f^{-1} (dz)^2\bigr]
   ,
\label {eq:metric}
\end {equation}
where $z$ is the coordinate $x^\five$ of the fifth dimension,
$\Rads$ is the radius of the 5-sphere (which will drop out of final results),
and
\begin {equation}
   f \equiv 1 - \frac{z^4}{\zh^4} \,.
\end {equation}
The boundary is at $z{=}0$, and the horizon is at
\begin {equation}
   \zh = \frac{1}{\pi T} \,.
\label {eq:zh}
\end {equation}
We will not need to worry about the details of regularizing the location
of the boundary in this work.


\subsection {Review of \boldmath$\lambda{=}\infty$ calculation}
\label {sec:geodesic}

Fig.\ \ref{fig:Wdecay} gave a cartoon picture of how we create our
``jets'' in the plasma.  Readers may find a precise description
of the field theory problem
in any of the previous papers \cite{adsjet,adsjet2,jj,R4} utilizing this
method, but we will not need those details here.  Suffice it to say
that, in the gravity description, the boundary is perturbed in a
localized region of space-time in such a way as to
create an excitation with large 4-momentum $q^\mu \simeq (E,0,0,E)$
and a relatively small amount of time-like 4-virtuality $-q^2 \ll E^2$.
The response of the system is then tracked at late times to see
where the excitation comes to a stop.

The supergravity field that is excited depends on the nature
of the source for the jet and is the supergravity field dual to the
vertex operator in fig.\ \ref{fig:Wdecay}.  For example, if we
study the decay of a slightly off-shell graviton in
the 4-dimensional quantum field theory, then the relevant supergravity
excitation is in the 5-dimensional gravitational field;
if we were to study the decay of a gauge boson weakly coupled to
R charge in the 4-dimensional quantum field theory, then the
supergravity excitation would be in a corresponding
5-dimensional gauge field; and so forth.

As mentioned in the introduction,
there is a nice simplification to the $\lambda{=}\infty$
analysis of
this problem \cite{adsjet2}
when $\lstop \ll \lmax$.  On the gravity side,
the wavepacket's motion is then the same (up to parametrically small
corrections) as that of a 5-dimensional ``particle'' traveling in
the AdS$_5$-Schwarzschild geometry, and so it follows a geodesic whose
trajectory is easily calculated in terms of the 4-momentum $q_\mu$ of
the excitation.  The exact geodesic depends on the mass of the 5-dimensional
``particle'' and so on the mass $m$ of the 5-dimensional supergravity
field that we have excited, but this mass (if any) may be ignored in the
high energy limit.%
\footnote{
  The relevant 5-dimensional supergravity field is the one dual to the
  operator used to create the ``jet'' excitation in ${\cal N}{=}4$
  SYM.  The mass $m$ of the supergravity field is determined by the
  conformal dimension $\Delta$
  of that operator, e.g.\ $(\Rads m)^2 = \Delta(\Delta-d)$
  for scalar operators
  \cite{Witten},
  where $d{=}4$.  We emphasize that $m$ is
  a mass in the 5-dimensional supergravity theory and has nothing to do with
  ``mass'' of a jet from the point of view of the 3+1 dimensional
  ${\cal N}{=}4$ SYM quantum field theory.
}

As a result, attention may be restricted to null geodesics, which
are given by
\begin {equation}
   x^\mu(x^\five) = \int \sqrt{g_{\five\five}} \, dx^\five \>
      \frac{ g^{\mu\nu} q_\nu }
           { (-q_\alpha g^{\alpha\beta} q_\beta)^{1/2} }
   \,.
\label {eq:geodesic}
\end {equation}
Using the metric (\ref{eq:metric}),
the $\lambda{=}\infty$ stopping
distance is then found to be \cite{adsjet2}
\begin {equation}
  \ell_{\rm stop} \simeq
  \int_0^{\zh} dz \> \frac{|\q|}{\sqrt{-q^2 + \frac{z^4}{\zh^4} |\q|^2}}
  \simeq
  \frac{\Gamma^2(\frac14)}{(4\pi)^{1/2}} \left( \frac{E^2}{-q^2} \right)^{1/4}
  \frac{1}{2\pi T}
  \,.
\label {eq:stop}
\end {equation}
An important feature of the integral in (\ref{eq:stop}) is that it
is dominated by small values of $z$, of order
\begin {equation}
   z_\star
   \equiv \zh \left( \frac{-q^2}{|\q|^2} \right)^{1/4}
   \simeq \zh \left( \frac{-q^2}{E^2} \right)^{1/4}
   \ll \zh .
\label {eq:zstar}
\end {equation}
$z_\star$ corresponds to the parametric scale $x^\five_\star$ in
fig.\ \ref{fig:fall}, where the trajectory turns over and
beyond which 3-space motion rapidly slows to a stop.
The stopping distance is determined by the behavior
of the trajectory at $z \sim z_\star$.


\section {Discussion of \boldmath$1/\sqrt{\lambda}$ expansion}
\label {sec:R4review}

\begin {figure}
\begin {center}
  \includegraphics[scale=0.5]{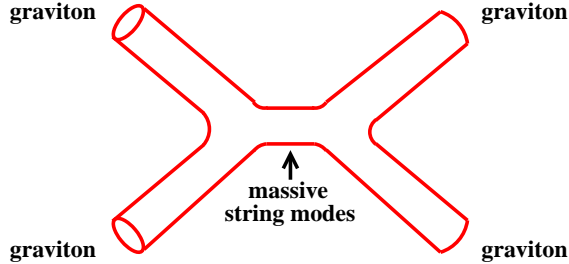}
  \caption{
     \label{fig:string4}
     A picture of massive string mode corrections to graviton-graviton
     scattering that are accounted for by the $D^m R^4$ corrections
     to the effective supergravity action.
  }
\end {center}
\end {figure}

We now want to consider corrections to the $\lambda{=}\infty$ results
for the stopping distance in the case $\Nc = \infty$.  First, we take
a moment to review the generic story of $1/\sqrt\lambda$ corrections in
the AdS/CFT correspondence, which relates
\cite{MAGOO}
\begin {equation}
  g_{\rm string} = \frac{\lambda}{4\pi\Nc} = \frac{g_{\rm YM}^2}{4\pi} \,,
\end {equation}
\begin {equation}
   \frac{1}{2\pi\tension \Rads^2} = \frac{\alpha'}{\Rads^2} = \lambda^{-1/2}
   \equiv (g_{\rm YM}^2 N_{\rm c})^{-1/2} ,
\end {equation}
where $g_{\rm string}$ is the string loop expansion parameter.
The string tension $\tension$ sets the mass scale
for massive string excitations, and so $\alpha'\to 0$ corresponds to taking the
scale for massive string excitations to infinity.
For $\lambda{=}\infty$, the strongly-coupled 4-dimensional quantum
field theory is therefore dual to the infrared limit
of the 10-dimensional string theory, namely supergravity, in the
appropriate background.  For large but finite $\lambda$, massive
string modes are not completely ignorable, and the effective
supergravity theory of the massless modes gets corrections, in
the form of higher-dimensional terms in its action, from integrating
out the effects of the massive modes.  Schematically, the
effective supergravity Lagrangian becomes%
\footnote{
  The precise details of which terms $D^m R^n$ appear independently
  in (\ref{eq:Leff})
  and which do not will not matter to our discussion, but Table I
  of ref.\ \cite{Stieberger} gives a nice summary of what's currently
  known at tree level (i.e.\ corresponding to $\Nc{=}\infty$).
}
\begin {multline}
   {\cal L} ~~\sim~~
   R
   ~~+~~
   \bigr[ \alpha'^3 R^4 + \alpha'^5 D^4 R^4 + \alpha'^6 D^6 R^4 + \cdots \bigl]
\\
   ~~+~~
   \bigr[ \alpha'^5 D^2 R^5 + \alpha'^6 D^4 R^5 + \alpha'^7 D^6 R^5
          + \cdots \bigl]
   ~~+~~ \cdots ,
\label {eq:Leff}
\end {multline}
where we have focused just on the gravitational fields for simplicity.
$R$ represents factors of the Riemann tensor, and we have not shown
numerical coefficients or how the indices contract.
For $\Nc{=}\infty$, there are no loop effects ($g_{\rm string}{=}0$),
and accounting for the massive string modes in the effective theory
is analogous to replacing the effects of the W boson by the
Fermi 4-point interaction in electroweak theory.  So, for example,
the $R^4$ terms in  (\ref{eq:Leff}) are calculated from
string amplitudes for $2\to 2$ graviton scattering and, crudely
speaking, they correspond to processes which involve intermediate
massive string states, as depicted schematically in fig.\ \ref{fig:string4}.
(A more accurate statement would be that they correspond to the full
string amplitude for graviton-graviton scattering
minus the sum of the $s$, $t$, and $u$-channel graviton
exchange diagrams that one would calculate in the $\alpha'{=}0$
supergravity theory.)  The $R^5$ terms similarly account for corrections
to the 5-point graviton interaction, and so forth.

\begin {figure}
\begin {center}
  \includegraphics[scale=0.40]{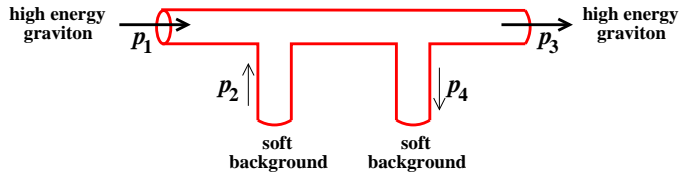}
  \caption{
     \label{fig:string4kin}
     A high-energy graviton, depicted as a string loop, interacting twice
     with the AdS$_5$-Schwarzschild background gravitational field.
  }
\end {center}
\end {figure}

In our application, we are interested in the evolution of a high-energy
excitation propagating through the soft AdS$_5$-Schwarzschild background.
For simplicity of presentation, we will focus on the case where the
excitation is in the 5-dimensional gravitational fields, though our
conclusions will not be sensitive to this assumption.
A classical gravitational excitation can also be thought of as a coherent
configuration of gravitons, and the relevant string scattering amplitudes
are those where two of the external lines are the incoming and outgoing
high-energy gravitons and the others are the soft background field.
So, for a 4-point scattering amplitude such as fig.\ \ref{fig:string4},
the relevant kinematic limit is that depicted in
fig.\ \ref{fig:string4kin}.
With the notation used in that figure, the high-energy
limit corresponds to potentially large $s=-(p_1+p_2)_I (p_1+p_2)^I$ but small
$t=-(p_1-p_3)_I (p_1-p_3)^I$.  [Here and throughout we may think of the
$p$'s as
5-dimensional momenta in AdS$_5$-Schwarzschild
rather than 10-dimensional momenta in (AdS$_5$-Schwarzschild)$\times S^5$
because in our problem there is no interesting dynamics associated with
the 5-sphere $S^5$.]
As discussed in ref.\ \cite{R4}, the $D^m R^4$ terms in (\ref{eq:Leff})
all become equally important in the jet stopping problem when this
5-dimensional $\sqrt{s}$ becomes large enough at $z \sim z_\star$
to excite massive string
modes in fig.\ \ref{fig:string4kin}.  The string mass scale is of
order $1/\sqrt{\alpha'}$, and this condition
\begin {equation}
  \sqrt{s_{\mbox{\scriptsize (5-dim)}^{\vphantom{2}}}} \gtrsim \frac{1}{\sqrt{\alpha'}}
\label {eq:Mthreshold}
\end {equation}
is shown in ref.\ \cite{R4} to be the same, in the context of the
jet stopping problem, as the condition
\begin {equation}
  \lstop \lesssim \lambda^{-1/6} \lmax ,
\end {equation}
which is the problematic case (\ref{eq:fun}) highlighted in our
introduction.  In this region, massive string states in the
intermediate state in fig.\ \ref{fig:string4kin} are kinematically
accessible and cannot be ignored.

\begin {figure}
\begin {center}
  \includegraphics[scale=0.40]{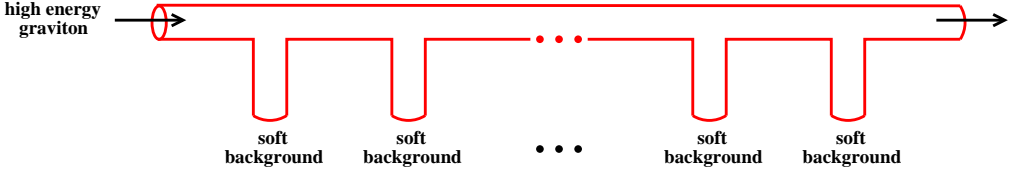}
  \caption{
     \label{fig:stringmany}
     Like fig.\ \ref{fig:string4kin} but with many interactions with
     the soft background field.
  }
\end {center}
\end {figure}

As the high-energy excitation falls from the boundary to the horizon,
as in fig.\ \ref{fig:fall}, it does not just interact with the
background field once or twice but does so over and over again, as
depicted in fig.\ \ref{fig:stringmany}.  If the massive string states
are kinematically accessible as in (\ref{eq:Mthreshold}), then they
cannot be neglected in any of the internal lines, which means in the
effective theory language of (\ref{eq:Leff}) that all $D^m R^n$ terms
will also become important.  This is just what happens at the
$\lstop \sim \lambda^{-1/6} \lmax$ point in fig.\ \ref{fig:R4}, where
all the corrections become the same size, corresponding to the
threshold $\sqrt{s_{\mbox{\scriptsize (5-dim)}^{\vphantom{2}}}} \sim
1/\sqrt{\alpha'}$.  Also, note that if massive string modes are
kinematically accessible for intermediate states, then they are also
accessible as final states.  (In our problem, however, there is never
really an ultimate ``final'' asymptotic state of the excited graviton
because the excited graviton falls into the black brane.)

Interaction with a pure AdS$_5$ background does not produce these
massive string mode excitations.  On a technical level, the $R^4$ interactions
really involve the Weyl tensor (the traceless part of the Riemann
tensor), which vanishes for AdS$_5$ but not for AdS$_5$-Schwarzschild.
So it is only the gravitational effect due to the presence of the
black brane that contributes to massive string mode excitation.
As a result, the effects of excited string modes
are negligible at the boundary and become
stronger as one moves away from it (and so closer to the black brane).
At some distance from the boundary which we will review later, the
gravitational effects of the black brane become strong enough that
(\ref{eq:Mthreshold}) is satisfied, which is when string modes
may first be excited.

From the point of view of an effective theory (\ref{eq:Leff}) of
gravitons, having all the correction terms become the same size
(or worse), seems like an hopeless disaster for the purpose of
computations.  However, the picture of fig.\ \ref{fig:stringmany}
suggests a different tack.  What is happening is that the 10-dimensional
gravitons which make up the classical
excitation are really tiny (quantum) loops of
string which are getting their internal string degrees of freedom
excited as they fall in the background gravitational field.
Specifically, internal degrees of freedom of a small object are
affected by gravitational {\it tidal}\/ forces, which try to compress
the object in some directions and stretch it in others.
In any case, consider the fate of a single graviton as depicted by
fig.\ \ref{fig:stringmany}: a high-momentum object moves through a
soft background field.  Various authors have previously studied
applications of the eikonal approximation to string scattering
\cite{eikonal,Veneziano}.
The upshot, as reviewed below, is that
fig.\ \ref{fig:stringmany} may be replaced by the evolution of
a single string quantized in the classical background field.

\begin {figure}
\begin {center}
  \includegraphics[scale=0.45]{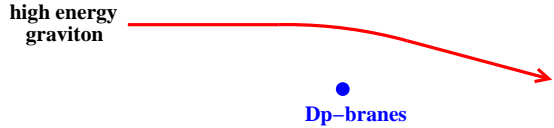}
  \caption{
     \label{fig:Dpscatt}
     A high energy massless string mode, such as a graviton, deflected
     by the gravitational field sourced by a stack of Dp-branes.  The
     plane of the figure is a plane orthogonal to the Dp-branes.
     (So, for instance, a D1-brane could be visualized as a line extending
     out of the page.)
  }
\end {center}
\end {figure}

\begin {figure}
\begin {center}
  \includegraphics[scale=0.45]{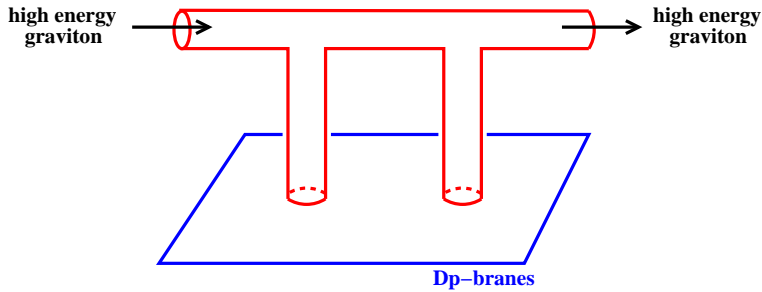}
  \caption{
     \label{fig:Dpscatt2}
     The 2nd-order scattering amplitude for a graviton to elastically
     scatter from a stack of Dp-branes, calculated as a string
     scattering amplitude with two connections to the Dp-branes.
     This is only a topological picture: As described in
     fig.\ \ref{fig:Dpscatt}, the motion of the graviton is orthogonal to
     the Dp-branes in the problem studied by
     D'Appollonio et al.\ \cite{Veneziano}.
  }
\end {center}
\end {figure}

One might simply assert that the right thing to do in the eikonal
limit is to quantize
a single string in the classical background field, but there is
a very nice paper by D'Appollonio, Di Vecchia, Russo, and Veneziano
\cite{Veneziano} that explicitly checks this in a closely related context.
As the source of their gravitational field, they take a stack of
$N$ coincident Dp-branes in otherwise-flat 10-dimensional space-time.
They then probe this gravitational field by scattering
a high momentum, massless closed string (such as a graviton) from it.
The geometry of the situation is depicted in fig.\ \ref{fig:Dpscatt},
which shows the probe particle moving in two of the asymptotically-flat
spatial
directions perpendicular to the Dp-branes.  The particle is deflected
by the gravitational field of the Dp-branes.  (If these were D0-branes
in 4-dimensional space-time, this could be a picture of a particle
deflected by the gravitational field of the Sun.)  Their problem
is slightly different from our problem in that their particle eventually
escapes back to infinity, given the geometry of their setup,
but never mind that.  They further assume that the impact parameter
$b$ in fig.\ \ref{fig:Dpscatt} is large enough that $bE \gg 1$.
This is the assumption that the background field experienced by the
particle is soft, its momentum components $\sim 1/b$ small compared to
the particle's energy.  After discussing the eikonal approximation more
generally, they then make the following check.  Consider the elastic
amplitude for the string loop to interact exactly twice with
the Dp-branes as it flies by and to emerge in the same massless state
at the end.  They perform this calculation in two different ways.
One way is
to do the full string calculation in the presence of the
D-branes, as depicted in fig.\ \ref{fig:Dpscatt2}.
The other way is to simply quantize a single string loop in the classical
gravitational background caused by the D-branes by taking the Penrose
limit of the metric near the string trajectory and quantizing the string in the
resulting pp-wave background.  Then they calculate to
second order in that background.  In the eikonal limit, they verify that
they get exactly the same result with either method.
The calculated probability
for the string to remain in its massless state drops rapidly below 1
once the kinematic threshold for exciting internal string
modes is exceeded.

With this reassurance, we now turn to taking the Penrose limit and
quantizing the string in our own problem.


\section {The Penrose Limit}
\label {sec:Penrose}

We begin by taking the Penrose limit to describe a narrow region
around the null geodesic (\ref{eq:geodesic}),
depicted earlier in
fig.\ \ref{fig:fallpp}.%
\footnote{
  Penrose limits have previously been studied in AdS$_5$-Schwarzschild
  by Pando Zayas and Sonnenschein \cite{LeoPenrose},
  but the null geodesic studied was different.
  Their geodesic fell straight toward the horizon in AdS$_5$-Schwarzschild,
  corresponding to $\q{=}0$ in our problem rather than $|\q| \simeq E$.
  Their geodesics also have non-trivial motion on the 5-sphere $S^5$.
  In our application,
  no dynamical evolution of the $S^5$ degrees of freedom takes place:
  the $S^5$ internal modes of the string stay in their ground state,
  and the $S^5$ zero modes of the string simply remain in a
  quantum state given by the
  $S^5$ harmonic of the supergravity field of interest.
  Equivalently, our worldsheet vacuum has $S^5$ conserved charges that
  are equal to those of the $S^5$ harmonic of the
  supergravity field of interest.
}
For an overview of taking Penrose limits, see
Refs.\ \cite{Blau,BlauLecture,BlauFermi}.

The null geodesic (\ref{eq:geodesic}) can be written as
\begin {subequations}
\label{eq:geodesic2}
\begin {equation}
   dx^\mu = \frac{g^{\mu\nu} q_\nu}{\omega} \> du ,
\end {equation}
where $u$ is an affine parameter for the geodesic determined by%
\footnote{
  This $u$ should not be confused with the coordinate
  $u \equiv (z/\zh)^2$ used in earlier work
  by some of the authors \cite{adsjet,adsjet2}.
}
\begin {equation}
   du = \omega \,
      \frac{ \sqrt{g_{\five\five}} \, dx^\five}
           { (-q_\alpha g^{\alpha\beta} q_\beta)^{1/2} }
   \,,
\label {eq:du}
\end {equation}
\end {subequations}
which can be integrated to give $u$ as a function of $x^\five$.
The normalization of $u$ is just convention, and the canceling factors
of $\omega=-q_\zero$ in (\ref{eq:geodesic2}) have just been chosen to give
$u$ dimensions of length.
In the metric (\ref{eq:metric}), $u$ is given by
\begin {equation}
   u = \Rads^2 \int \frac{dz}{z^2(1 - f |\q|^2/\omega^2)^{1/2}}
   \,,
\label {eq:uz0}
\end {equation}
which is linearly divergent as $z \to 0$.
Our convention will be to define $u$ as running from $u(z{=}0) = -\infty$ at
the boundary to $u(z{=}\infty) = 0$ at the black brane singularity, and so
\begin {equation}
   u(z) = - \Rads^2 \int_z^{\infty} \frac{dz}{z^2(1 - f |\q|^2/\omega^2)^{1/2}}
   \,.
\label {eq:uformula}
\end {equation}
The important thing to remember in what follows
is that {\it late}\/ times correspond to {\it small}\/ negative
values of $u$ (which we will later also call $\tau$).
Also, though it will be convenient to have defined $u=0$ to be
at the singularity $z = \infty$, all of the physics we discuss will
only depend
on what happens outside of the horizon, $z < \zh$.

We'll refer to our reference geodesic as $\bar x^\mu(x^\five)$, which we
choose to start at the origin $\bar x^\mu = 0$ on the boundary.  Now measure
4-positions in AdS$_5$-Schwarzschild relative to this geodesic
by defining
\begin {equation}
   \Delta x^\mu \equiv x^\mu - \bar x^\mu(x^\five) .
\label{eq:deltax}
\end {equation}
Take $\q$ to be in the $x^\three$ direction.  Changing
coordinates from $x^\five$ and $\Delta x^\zero$ to $u{=}u(x^\five)$ and
\begin {equation}
   v \equiv \frac{q_\mu \> \Delta x^\mu}{\omega}
   = - \Delta x^\zero + \frac{|\q|}{\omega} \, \Delta x^\three
\label {eq:v}
\end {equation}
puts the AdS$_5$-Schwarzschild metric (\ref{eq:metric}) into the
form
\begin {multline}
   (ds)^2 =
   2 \, du \, dv
   + \frac{\Rads^2}{z^2} \Biggl[
       (dx^\one)^2 + (dx^\two)^2
       + \frac{(\omega^2 - f |\q|^2)}{\omega^2} \, (d\Delta x^\three)^2 
\\
       + 2 f \, \frac{|\q|}{\omega} \, dv \, d\Delta x^\three
       - f \, (dv)^2
     \Biggr] ,
\label {eq:metricuv}
\end {multline}
where $f$ is now implicitly a function of $u$.
The Penrose limit consists of keeping only the terms in the metric
that would dominate after a scaling of coordinates
\begin {equation}
   u \to u,
   \quad
   v \to \gamma^{-2} \, v
   \quad
   x^i \to \gamma^{-1} x^i
\label{eq:scaling}
\end {equation}
for very large $\gamma$.
This is analogous to making what would be a very large boost
($u \to \gamma u$, $v \to \gamma^{-1} v$) in flat space, and so
looking at physics close to the light cone, followed by rescaling
all coordinates by a factor of $\gamma^{-1}$.
For (\ref{eq:metricuv}), the resulting limit is
\begin {equation}
   (ds)^2_{\rm pp} =
   2 \, du \, dv
   + \frac{\Rads^2}{z^2} \Biggl[
       (dx^\one)^2 + (dx^\two)^2
       + \frac{(\omega^2 - f |\q|^2)}{\omega^2} \, (d\Delta x^\three)^2 
     \Biggr] ,
\label {eq:metricpp0}
\end {equation}
which is a particular example of a pp-wave metric.
When we are done with our analysis, we will go back and check that
this approximation is justified.
The coordinates used in (\ref{eq:metricpp0}) are known as Rosen
coordinates.

The metric (\ref{eq:metricpp0}) has the schematic form
\begin {equation}
   (ds)^2 = 2 \, du \, dv + \sum_i \kappa_i(u) \, (dy_i)^2 .
\label {eq:metricyhat}
\end {equation}
It is useful to normalize the last term by switching
to Brinkmann coordinates
\begin {equation}
  \hat y_i \equiv y_i \sqrt{\kappa_i(u)}
\label {eq:yhat}
\end {equation}
and
\begin {equation}
  \hat v \equiv
  v - \tfrac12 \sum_i \partial_u(\ln \sqrt{\kappa_i}) \, \hat y_i^2 
\label {eq:vhat}
\end {equation}
to give
\begin {equation}
  (ds)^2 =
  2 \, du \, d\hat v
  + \sum_i (d\hat y_i)^2
  + \Bigl(\sum_i \frac{\partial_u^2 \sqrt{\kappa_i}}{\sqrt{\kappa_i}}
                  \, \hat y_i^2 \Bigr) (du)^2
  \,.
\end {equation}
In our case, using (\ref{eq:uz0}) to rewrite
\begin {equation}
  \partial_u =
  \frac{z^2}{\Rads^2} \, \Bigl(1-f\frac{|\q|^2}{\omega^2}\Bigr)^{1/2} \partial_z ,
\label {eq:partialu}
\end {equation}
the metric in Brinkmann coordinates is
\begin {equation}
  ds^2_{\rm pp} =  2 \, du \, d\hat v
  + (d\hat x^\one)^2 + (d\hat x^\two)^2 + (d\Delta\hat x^\three)^2
   + {\cal G}(u,\hat x^\one, \hat x^\two, \Delta\hat x^\three) \,
     (du)^2
\label {eq:ppmetrichat}
\end {equation}
with
\begin {subequations}
\label {eq:Gdefs}
\begin {equation}
  {\cal G}(u,\hat x^\one,\hat x^\two,\Delta\hat x^\three)
  = {\cal G}_1(u) \, \bigl[(\hat x^\one)^2 + (\hat x^\two)^2\bigr]
           + {\cal G}_3(u) \, (\Delta\hat x^\three)^2 ,
\end {equation}
\begin {equation}
  {\cal G}_1(u) = {\cal G}_2(u)
  = \frac{\partial_u^2(z^{-1})}{z^{-1}}
  = \frac{z^3 f' |\q|^2}{2 \Rads^4\omega^2}
  = - 2 \, \frac{z^6 |\q|^2}{\zh^4 \Rads^4\omega^2}
  \simeq - 2 \, \frac{z^6}{\zh^4 \Rads^4}
  \,,
\label {eq:calG12}
\end {equation}
\begin {equation}
  {\cal G}_3(u)
  = \frac{\partial_u^2\bigl[z^{-1}(\omega^2-f|\q|^2)^{1/2}\bigr]}
          {z^{-1}(\omega^2-f|\q|^2\bigr)^{1/2}}
  = \frac{z^3 (f'-zf'') |\q|^2}{2 \Rads^4 \omega^2}
  = 4 \frac{z^6 |\q|^2}{\zh^4 \Rads^4 \omega^2}
  \simeq 4 \frac{z^6}{\zh^4 \Rads^4}
  \,.
\label {eq:calG3}
\end {equation}
\end {subequations}
Here, primes denote derivatives with respect to $z$, and $z{=}z(u)$
is implicitly a function of $u$, determined by inverting
(\ref{eq:uformula}).
The metric (\ref{eq:ppmetrichat}) would be flat if not
for the ${\cal G}_i$, which arise from tidal forces.
Note that these tidal terms vanish for null geodesics in
pure AdS ($f{=}1$, or equivalently $\zh{\to}\infty$), in agreement
with general arguments that the Penrose limit of AdS is flat Minkowski
space \cite{Blau}.
They also vanish in AdS$_5$-Schwarzschild if $\q{=}0$
(i.e. if the excitation fell straight down in the $x^\five$
direction as a function of time $x^\zero$).


\section {Quantizing our falling string loop}
\label {sec:quantize}

\subsection {Overview}

We may now follow other authors to quantize a string in a
pp-wave background \cite{Veneziano,PRT,HS,GPZS,dVS}.
The bosonic sector is described by a $\sigma$-model in the
pp-wave background metric:
\begin {equation}
   S =
   - \frac{1}{4\pi\alpha'} \int d\tau \> d\sigma
       \sqrt{-\gamma} \, \gamma^{\alpha\beta}
       (\partial_\alpha X^I)(\partial_\beta X^J) \,
       g_{IJ}(X) ,
\end {equation}
where $\gamma$ is the world-sheet metric and $X^I$ are the world-sheet
fields corresponding to the coordinates.
For the pp-wave space-time
metric (\ref{eq:ppmetrichat}), this takes the form
\begin {equation}
   S = S_0
   - \frac{1}{4\pi\alpha'} \int d\tau \> d\sigma
       \sqrt{-\gamma} \, \gamma^{\alpha\beta}
       (\partial_\alpha U)(\partial_\beta U) \,
       {\cal G}(U,\Delta \hat {\bm X}) ,
\label {eq:action}
\end {equation}
where $S_0$ is the Minkowski string action.
Identifying world-sheet time
$\tau$ with the affine parameter $u$ for the pp-wave space-time
metric (\ref{eq:ppmetrichat}) then gives a constraint equation
for $\partial_\alpha\hat V$ and gives the light-cone gauge Lagrangian
\begin {align}
   L &=
   \frac{p^u}{2} \int_0^{2\pi} \frac{d\sigma}{2\pi} \sum_i
       \left(
          \partial_\tau \Delta \hat X^i \, \partial_\tau \Delta \hat X^i
          - (\alpha'p^u)^{-2}
               \partial_\sigma \Delta \hat X^i \,
               \partial_\sigma \Delta \hat X^i
          + {\cal G}_i(\tau) \, \Delta \hat X^i \, \Delta\hat X^i
       \right)
\nonumber\\
   &=
   \frac{p^u}{2} \sum_i \sum_{n=-\infty}^{\infty}
       \left(
          \partial_\tau \Delta \hat{X^i_n}^* \, \partial_\tau \Delta \hat X^i_n
          - \omega_{i,n}^2(\tau) \,
            \Delta \hat{X^i_n}^* \, \Delta\hat X^i_n
       \right)
\label {eq:Lin}
\end {align}
for the $\Delta\hat X^i$, where $p^u = p_{\hat v} \simeq E$ and
$\Delta \hat X^i = \sum_n \Delta \hat X^i_n(\tau) \, e^{in\sigma}$ and
\begin {equation}
    \omega_{i,n}^2(\tau) \equiv
              \frac{n^2}{(\alpha' p^u)^2} - {\cal G}_i(\tau) .
\label {eq:win}
\end {equation}
(We have suppressed the fields corresponding to $S^5$
coordinates because they will play no role in our discussion.)
Note that we have chosen a convention where $\tau$ has units of
time and $\sigma$ is dimensionless.%
\footnote{
   A more conventional normalization might be to define a
   dimensionless $\tau$ as
   $\tau \equiv u/(\alpha' p^u)$ instead
   of our $\tau \equiv u$.  One may convert to this convention by
   everywhere replacing our $\tau$ by $\alpha' p^u \tau$.
   In our convention, the world-sheet metric is
   \[
      \gamma_{\alpha\beta} \propto
      \begin{pmatrix}
         -(\alpha' p^u)^{-1} & 0 \\
         0 & \alpha' p^u
       \end {pmatrix} .
    \]
}

Each mode $\Delta\hat X^i_n$ of the string is a time-dependent harmonic
oscillator problem with classical frequency $\omega_{i,n}(\tau)$.
The $\Delta\hat X^\one_n$ and $\Delta\hat X^\two_n$ modes are tidally
compressed as the string moves away from the
boundary since ${\cal G}_1 = {\cal G}_2$ is negative in
(\ref{eq:calG12}), so that the curvature $\omega_{1,n}(\tau)$
of the harmonic oscillator potential increases
with time $\tau{=}u$.
In contrast, the $\Delta\hat X^\three_n$ oscillators are
tidally stretched, since ${\cal G}_3$ is positive in (\ref{eq:calG3}).
We will focus on these $\Delta\hat X^\three$ degrees of freedom and
so will focus on
\begin {equation}
    \omega_{3,n}^2(\tau) =
              \frac{n^2}{(\alpha' p^u)^2} - {\cal G}_3(\tau) .
\label {eq:w3n}
\end {equation}
When the string gets far enough from the boundary (i.e.\ at
late enough times $\tau$), the ${\cal G}_\three$ term in
(\ref{eq:w3n}) becomes dominant and $\Delta \hat X^\three_n$
oscillators become unstable.  Physically, this is when tidal
forces come to dominate over string tension.
Using (\ref{eq:calG3}), this instability occurs when
$z \ge z_n$, where
\begin {equation}
  z_n
  \simeq \left( \frac{n \zh^2 \Rads^2}{2 \alpha' E} \right)^{1/3}
  = \lambda^{1/6} \left( \frac{n \zh^2}{2 E} \right)^{1/3}
  = n^{1/3} z_1 .
\end {equation}
The instability for the center-of-mass mode $n{=}0$ is not
particularly interesting: it has nothing to do with exciting internal
degrees of freedom of the string and just reflects the slight spread
of the falling wavepacket in fig.\ \ref{fig:fall} due to curvature
effects.  That is, it reflects physics already incorporated in earlier
work \cite{adsjet2} on our jet stopping problem
and has nothing to do with gravitons being
tidally stretched into classical strings.  So, in what
follows, we will focus on the $n\not=0$ modes, and the first
tidal instability kicks in at $z \simeq z_1$.

Recall that $z_\star$, defined in (\ref{eq:zstar}), characterizes
the scale where the $x^\three$ motion of the bulk excitation in
fig.\ \ref{fig:fall}
is coming to a stop (no significant motion for $z \gg z_\star$).
How does the instability scale $z_n$ above compare to $z_\star$?
Using (\ref{eq:stop}) and (\ref{eq:lmax}), the ratio is
\begin {equation}
  \frac{z_n}{z_\star}
  \sim \frac{n^{1/3} \lambda^{1/6} (\zh^2/E)^{1/3}}{\zh (-q^2/E^2)^{1/4}}
  \sim \frac{n^{1/3} \lstop}{\lambda^{-1/6} \lmax} \,.
\end {equation}
So the tidal instability kicks in before the stopping distance
is reached ($z_1 \lesssim z_\star$) precisely when we are in the
interesting regime $\lstop \lesssim \lambda^{-1/6} \lmax$ (\ref{eq:fun})
identified in earlier work as the case where stringy corrections
become important.  This is the case we focus on.  Correspondingly,
the modes which become tidally unstable for
$z \lesssim z_\star$ are $n \lesssim n_\star$ with
\begin {equation}
   n_\star \sim \left(\frac{\lstop}{\lambda^{-1/6} \lmax}\right)^{-3}
   .
\label {eq:nstar}
\end {equation}

Although the instability develops at $z = z_n$, we will see later
that the
modes $n \lesssim n_\star$ do not have time to stretch significantly
until $z \sim z_\star$.
This fact is suggested by the geodesic picture in fig.\ \ref{fig:boost}:
For $z\ll z_\star$, the impact of the black hole on the evolution in
fig.\ \ref{fig:boost}c is negligible, and so the evolution at those
times is well approximated by the pure AdS case of figs.\
\ref{fig:boost}a and b, for which the proper size of the string
remains constant.

Once a given mode becomes unstable, the quantum mechanics of that mode
will be somewhat analogous to a standard quantum mechanics thought
experiment: What is the longest time that an idealized pencil
can be balanced on its tip before it falls?  Because of the
uncertainty principle, the pencil cannot be started simultaneously
at rest and perfectly vertical, and so it must fall.
The pencil might be started in a Gaussian wavepacket chosen to
maximize the average fall time.  But it is clear that, once the
top of the pencil has fallen a macroscopic distance,
{\it classical}\/ mechanics
will suffice to describe its subsequent motion: at that time,
its position and momentum may, to excellent approximation, be considered as
simultaneously well defined.  For late enough
times $t_1$, the pencil's motion for $t > t_1$ is approximately
classical,
with the only effect of its initial wave packet at $t{=}0$
being to determine
a classical probability distribution for the pencil end's position at
$t_1$.  The corresponding momentum at $t_1$ is
given (to excellent approximation) by
the momentum the pencil would have picked up falling classically
to that position from vertical.

In our problem, we will see that, for the string modes
$n \lesssim n_\star$ of
interest, the analogous transformation from a quantum
description to a probability distribution for classical configurations
occurs when $z \gg z_\star$.


\subsection {Solution of the time-dependent harmonic oscillators}

\subsubsection {Basics}

The distinction between $\Delta X^i$ (the difference between $X^i$ and
the reference geodesic) and $X^i$ does not affect the $n \not= 0$
modes that are our focus.  Similarly for the normalized coordinates
$\Delta \hat X^i$.  So we will make our notation a
little less cumbersome and henceforth
write $\Delta\hat X^i_n$ as simply $\hat X^i_n$
(for $n\not=0$).

Each of the real degrees of freedom
$\sqrt2 \Re \hat X^i_n$ and
$\sqrt2 \Im \hat X^i_n$ in (\ref{eq:Lin}) have a harmonic oscillator
Lagrangian of the form
\begin {equation}
  L = \tfrac12 \, m \bigl( \dot\dof^2 - \omega^2(\tau) \, \dof^2 \bigr) ,
\label {eq:Lho}
\end {equation}
with the translation $m \to p^u \simeq E$ and
$\omega^2(\tau) \to \omega_{i,n}^2(\tau)$.
The squared frequency $\omega^2(\tau)$ starts at a non-zero
value $\omega^2(-\infty)$ and then changes with time $\tau$.
The quantum mechanical solution to such time-dependent
harmonic oscillator problems has a long history.
Useful explicit formulas for wave functions may be found in
ref.\ \cite{KimLee}, with applications
to strings in pp-wave backgrounds in refs.\ \cite{PRT,GPZS}.%
\footnote{
  The case (\ref{eq:psi}) of interest to us corresponds to
  the ground state cases ($\ell{=}0$ or $n{=}0$) of
  refs.\ \cite{PRT,GPZS,KimLee}, and our $\chi$ here corresponds to
  their $\chi^*$, $v^*$, or $u^*$ respectively.  However, there are
  various
  normalization issues in the formulas in all of these references,
  which all differ from us and from each other by overall powers
  of our $\chi^*/\chi$ in their expressions
  for the final wave function, though most differences are probably
  typographic errors.  $\chi^*/\chi$ is just a time-dependent
  but $\dof$-independent complex phase, and so such differences
  will not matter for the
  evaluation of an expectation value in one of these states.
  But, if one wants explicit solutions to the time-dependent Schr\"odinger
  equation, then the phase needs to be correct.  A simple test of any such
  result is to consider the case where $\omega(\tau)$ is constant
  and verify that $\psi_n(\dof,t)$ reproduces the correct time dependence
  $\exp\bigl[-i(n+\tfrac12)\omega \tau\bigr]$.  Accordingly, the
  $(\chi/\chi^*)^\ell$ in (5.37) of ref.\ \cite{PRT} should be
  $(\chi/\chi^*)^{\frac{\ell}{2}+\frac14}$, the 
  $[v^i/{v^i}^*]$ in (2.16) [(2.15) in the preprint]
  of ref.\ \cite{GPZS} should
  be $[v^i/{v^i}^*]^{\frac{\ell}{2}+\frac14}$, and
  the $(u/u^*)^{n/2}$ in (3.9) of ref.\ \cite{KimLee} should
  be $(u/u^*)^{\frac{n}{2}+\frac14}$ if one wants to keep track
  of the overall time-dependent phases and have results that
  solve the Schr\"odinger equation.
  {\it Note added:} Formulas with the correct phase may be found in
  ref.\ \cite{KimPage}.
}
In our case, the harmonic oscillators all start in their ground
state (the string state describing a graviton) at early
times ($\tau \to -\infty$) and so start with Gaussian wave functions.
For a time-dependent harmonic oscillator that starts as a Gaussian
$\psi(\dof) \propto \exp[-\dof^2/4\sigma^2]$ at some time $\tau_0$, one
may check that the Schr\"odinger equation
\begin {equation}
  i \dot\psi
  = \Bigl[ - \frac{1}{2m} \, \partial_\dof^2 
       + \frac12 \, m \, \omega^2(\tau) \, \dof^2 \Bigr] \psi
\end {equation}
is solved by
\begin {equation}
   \psi(\dof,\tau) \propto
   \frac{1}{\sqrt{\chi(t)}} \,
   \exp\left[
        \frac{i}{2} \, \frac{\dot\chi(t)}{\chi(t)} \, m \dof^2
   \right] ,
\label {eq:psi}
\end {equation}
where the complex-valued function
$\chi(\tau)$ satisfies the classical equation of motion
\begin {equation}
   \ddot \chi = - \omega^2(\tau) \, \chi
\label {eq:chi}
\end {equation}
with initial conditions
\begin {equation}
   \chi(\tau_0) = 1,
   \qquad
   \dot\chi(\tau_0) = \frac{i}{2m\sigma^2} \,.
\end {equation}
In our case, where we start in the early-time ground state,
that's 
\begin {equation}
   \chi(\tau_0) = 1,
   \qquad
   \dot\chi(\tau_0) = i \, \omega(-\infty)
\label {eq:chii}
\end {equation}
with $\tau_0 \to -\infty$.

It will be useful to have an expression for the corresponding
probability distribution $|\psi(\dof,\tau)|^2$ for $\dof$.
From (\ref{eq:psi}), this is just a Gaussian distribution
\begin {equation}
   \operatorname{Prob}(\dof,\tau) =
   \frac{e^{-\dof^2/2\dof_{\rm rms}^2(\tau)}}{(2\pi)^{1/2} \, \dof_{\rm rms}(\tau)}
\end {equation}
with width
\begin {equation}
   \dof_{\rm rms} = \left[ 2m \Im \frac{\dot\chi}{\chi} \right]^{-1/2}
   = \left[ \frac{m(\chi^*\dot\chi-\dot\chi^*\chi)}{i|\chi|^2} \right]^{-1/2}
   .
\end {equation}
But $\chi^*\dot\chi-\dot\chi^*\chi$ is a Wronskian of the two solutions
$\chi$ and $\chi^*$ of (\ref{eq:chi}) and so is time-independent and
may be evaluated at $\tau{=}\tau_0$ using (\ref{eq:chii}), giving
\begin {equation}
   \dof_{\rm rms}(\tau)
   = \frac{|\chi(\tau)|}{\sqrt{2m\,\omega(-\infty)}}
   = |\chi(\tau)| \, \dof_{\rm rms}(-\infty) .
\label {eq:xrms}
\end {equation}
Using (\ref{eq:w3n}),
$x_{\rm rms}(-\infty)$ corresponds to $\sqrt{\alpha'/2n}$ in our
application.

Our remaining task is to solve the classical equation of motion
(\ref{eq:chi}) for $\chi$.  For the case of interest $\hat X^\three_n$,
using
(\ref{eq:zstar}), (\ref{eq:partialu}), (\ref{eq:calG3}),
(\ref{eq:w3n}), and $\omega \simeq |\q| \simeq E$,
the $\chi$ equation may be put into the form
\begin {subequations}
\label {eq:diffeqs}
\begin {align}
   \frac{d^2\chi}{d\bar\tau^2}
   &= -4(\xi^6 - \bar z^6) \chi ,
\label {eq:chixi}
\\
   \frac{d\bar z}{d\bar\tau} \> &= \bar z^2 (1+\bar z^4)^{1/2} ,
\label {eq:zbar}
\end {align}
\end {subequations}
where
\begin {align}
   \bar z &\equiv \frac{z}{z_\star} ,
\\
   \bar\tau &\equiv \frac{z_\star^3}{\zh^2 \Rads^2} \, \tau ,
\\
   \xi &= \xi_n \equiv n^{1/3} \xi_1 \equiv
   \left(\frac{n}{2}\right)^{1/3}
     \frac{\zh^2/z_\star}{\lambda^{-1/6} \zh^{4/3} E^{1/3}} .
\label {eq:xidef}
\end {align}
In these variables, the initial conditions
(\ref{eq:chii}) on $\chi$ are
\begin {equation}
   \chi(\bar\tau_0) = 1,
   \qquad
   \frac{d\chi}{d\bar\tau}(\bar\tau_0) = 2i\xi^3 \,.
\label {eq:chiinit}
\end {equation}
Note from (\ref{eq:lmax}) and (\ref{eq:stop}) that, parametrically,
\begin {equation}
  \xi_1 \sim \frac{\lstop}{\lambda^{-1/6}\lmax} ,
\end {equation}
and so the smallness of $\xi_1$ is a specific
measure of how far we are into the interesting regime (\ref{eq:fun})
of $\lstop \lesssim \lambda^{-1/6}\lmax$.
The modes $n \lesssim n_\star$ of interest to us correspond to
$\xi_n \lesssim 1$.

Using the above equations, one may now check
our earlier claim that
the string does not stretch significantly in proper size at
early times $z \ll z_\star$ ($\bar z \ll 1$).%
\footnote{
  The argument is a distraction and so relegated to this footnote.
  For $z_n \ll z \ll z_\star$ ($\xi \ll \bar z \ll 1$), one may
  ignore the $\xi^6$ term in (\ref{eq:chixi}) and treat $\bar z$
  as small.  In this case the two independent solutions are
  approximately ${\cal A} \simeq (\xi/\bar z)(1 + \tfrac12 \, \bar z^4)$ and
  ${\cal B} \simeq 1 + \tfrac15 \, \bar z^4$,
  with
  $\partial_{\bar\tau} \simeq \bar z^2 (1 + \tfrac12 \bar z^4) \partial_{\bar z}$.
  (See Appendix \ref{app:PenroseCheckString} for
  a more general discussion of $z \gg z_n$ solutions, for which the ones
  here are the small $\bar z$ limit.)
  To match to the oscillating
  solution for $\chi$ for $\bar z \ll \xi$ determined by (\ref{eq:chiinit}),
  the superposition of ${\cal A}$ and ${\cal B}$ for $\bar z \gg \xi$
  must be $\chi \simeq O(1) \, {\cal A} + O(1) \, {\cal B}$, where
  $O(1)$ denotes coefficients with magnitude of order 1.
  We then see that $\chi$ is approximately constant over the time
  period $\xi \ll \bar z \ll 1$.
}  
But we are more interested
in what the string does at late times, on which we now focus.


\subsubsection {Late-time behavior}
\label {sec:late}

For $z \gg z_\star$ (which is $\bar z \gg 1$), the $\bar z$
equation (\ref{eq:zbar}) gives
\begin {equation}
   \bar z = (-3 \bar\tau)^{-1/3} ,
\label {eq:zbarlate}
\end {equation}
remembering that our convention is that $\tau$ is negative and that
$\tau(z{=}\infty)=0$.  Note from (\ref{eq:zbarlate})
that $-\bar\tau$ is very small at the horizon $z = \zh$,
where
$-\bar\tau \sim (\zh/z_\star)^{-3} \sim (-q^2/E^2)^{3/4} \sim (\lstop T)^{-3}$.
Throughout this paper we assume that $-q^2 \ll E^2$ and so
$\lstop \gg T^{-1}$.

Plugging (\ref{eq:zbarlate}) into the $\chi$ equation
(\ref{eq:chixi}) yields late-time ($\bar z \gg 1$) solutions
$\chi \propto (-\bar\tau)^{-1/3}$ and $\chi \propto (-\bar\tau)^{4/3}$.
The dominant solution will be
\begin {equation}
   \chi \propto (-\bar\tau)^{-1/3} .
\label {eq:chilate}
\end {equation}

Though $\chi$ is a complex-valued function whose purpose is to
track the evolution of the wavepacket, exactly the same arguments as
above give that a classical trajectory would have late-time
behavior $\dof \propto (-\bar\tau)^{-1/3}$.  That means that
$\dot\dof \propto (-\bar\tau)^{-4/3}$, and so $\dof$ and $\dot\dof$ both
become large at late times, justifying a classical description at
late times.%
\footnote{
   For example, at late times the exponential in the wavepacket
   (\ref{eq:psi}) becomes $\exp[i S]$, where $S \propto \dof^2/(-\tau)$.
   The WKB condition $|\partial_\dof^2 S| \ll (\partial_\dof S)^2$ is
   satisfied as $\tau \to 0$ for $\dof \propto (-\tau)^{-1/3}$.
   We will see shortly that the proportionality constant in
   (\ref{eq:chilate}) is of order 1 for the modes $n \lesssim n_\star$ of
   interest.  If one keeps track parametrically
   of all the proportionality constants in the exponential $\exp[i S]$,
   one finds more specifically
   that the WKB condition is satisfied when
   $-\bar\tau \ll 1$ (i.e.\ $\bar z \gg 1$).
}
The classical relation between the two is determined by
$\dof \propto (-\bar\tau)^{-1/3}$ to be
\begin {equation}
   \frac{\dot x}{x} \simeq \frac{1}{-3\bar\tau} \simeq \bar z^3 .
\label {eq:xdotclassical}
\end {equation}

\begin {figure}
\begin {center}
  \includegraphics[scale=0.5]{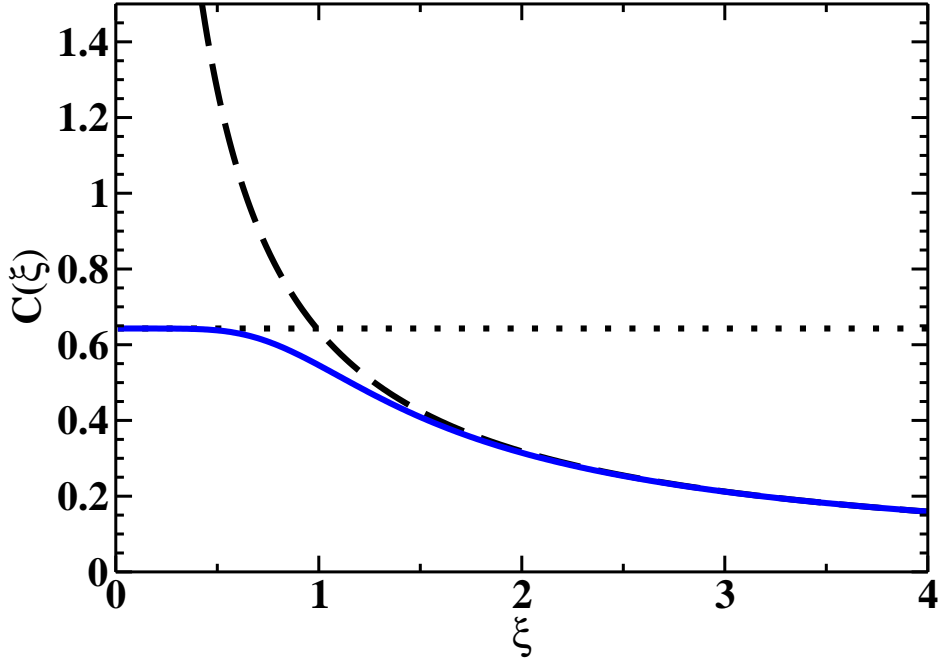}
  \caption{
     \label{fig:Cxi}
     The proportionality constant $C(\xi)$ in (\ref{eq:Cdef}), which
     determines the late-time width of the probability distribution
     for the amplitude of a string mode.  The sloping dashed curve
     shows the large-$\xi$ approximation (\ref{eq:Clargexi}).
  }
\end {center}
\end {figure}

We may extract the proportionality constant in the late-time behavior
(\ref{eq:chilate}) by solving (\ref{eq:diffeqs}) numerically with
initial conditions (\ref{eq:chiinit}) for $\chi$, matching the late
time behavior of the numerical solution to (\ref{eq:chilate}),
and repeating the calculation for earlier and earlier values of
$\tau_0$ in order to take the $\tau_0 \to -\infty$ limit.
Our result is that the late-time behavior is
\begin {equation}
   |\chi(\bar\tau)| \to \frac{C(\xi)}{(-\bar\tau)^{1/3}}
\label {eq:Cdef}
\end {equation}
with $C(\xi)$ given by fig.\ \ref{fig:Cxi}.%
\footnote{
  For numerical work, it is mildly convenient to eliminate $\bar\tau$
  and express all of the relevant equations solely in terms of
  $\bar z$, giving
  $\bar z^4 (1+\bar z^4) \chi'' + 2 \bar z^3(1+2\bar z^4) \chi'
   = -4(\xi^6-\bar z^6)\chi$
  and $\chi'(\bar z_0) = 2 i \xi^3/\bar z_0^2$
  (with $\bar z_0 \to 0$)
  and $|\chi(\bar z)| \to 3^{1/3} \, C(\xi) \, \bar z$
  (as $\bar z \to \infty$).
}
We show in
Appendix \ref{app:Cxibig} that the limiting behavior for
large $\xi$ is
\begin {subequations}
\label {eq:Cxilimits}
\begin {equation}
   C(\xi) \simeq \frac{\Gamma(\tfrac56)}{\pi^{1/2}\xi}
   \qquad
   \mbox{for}~\xi \gg 1 ,
\label {eq:Clargexi}
\end {equation}
shown as a dashed curve in the figure.
In the opposite limit of $\xi \ll 1$, our numerical results
approach a constant
\begin {equation}
   C(\xi) \simeq 0.6428
   \qquad
   \mbox{for}~\xi \ll 1.
\label {eq:C0}
\end {equation}
\end {subequations}

From (\ref{eq:w3n}), (\ref{eq:xrms}) and (\ref{eq:Cdef}),
and remembering that the analogs of $x$ are
$\sqrt{2} \Re X^i_n$ and $\sqrt{2} \Im X^i_n$,
the late time probability distribution of mode amplitudes
$\hat X^\three_n$ is given by a Gaussian with width
\begin {equation}
   \bigl|\hat X^\three_n\bigr|_{\rm rms}
   \simeq
   \frac{C(\xi_n)}{(-\bar\tau)^{1/3}}
     \left(\frac{\alpha'}{2n}\right)^{1/2}
   \qquad
   \mbox{for}~{-}\bar\tau \ll 1.
\end {equation}
Using (\ref{eq:zbarlate}),
that may be rewritten as
\begin {equation}
   \bigl|\hat X^\three_n\bigr|_{\rm rms}
   \simeq
   3^{1/3} \, C(\xi_n) \, \frac{z}{z_\star}
     \left(\frac{\alpha'}{2n}\right)^{1/2}
   \qquad
   \mbox{for}~ z \gg z_*,\xi_n.
\label {eq:Xhatrms}
\end {equation}
As in (\ref{eq:xdotclassical}), the corresponding
momenta in this classical regime are related to the mode
amplitudes $\hat X^\three_n$ by
\begin {equation}
   \partial_{\bar\tau} \hat X^\three_n
   \simeq \frac{\hat X^\three_n}{-3\bar\tau}
   \simeq \frac{z^3}{z_\star^3} \, \hat X^\three_n \,.
\end {equation}

Using (\ref{eq:metricpp0}) and (\ref{eq:yhat}), the conversion between
the normalized coordinate $\Delta \hat x^\three$ and the
displacement $\Delta x^\three$ from the reference geodesic in
Poincare coordinates is
\begin {equation}
   \Delta x^\three =
   \frac{z}{\Rads} \left( 1 - f \frac{|\q|^2}{\omega^2} \right)^{-1/2}
   \simeq
   \frac{z}{\Rads} \left( \frac{z_\star^4 + z^4}{\zh^4} \right)^{-1/2}
   \Delta\hat x^\three ,
\end {equation}
which is
\begin {equation}
   \Delta x^\three \simeq
   \frac{\zh^2}{z \Rads} \, \Delta\hat x^\three
   \qquad \mbox{for}~z \gg z_\star .
\end {equation}
So, from (\ref{eq:Xhatrms}),
the amplitudes of the stretched modes in the Poincare coordinate
system are
\begin {equation}
   \bigl|X^\three_n\bigr|_{\rm rms}
   \simeq
   3^{1/3} \, C(\xi_n) \, \frac{\zh^2}{z_\star \Rads}
     \left(\frac{\alpha'}{2n}\right)^{1/2}
   =
   \frac{3^{1/3} \, C(\xi_n) \, \zh^2}{(2n)^{1/2} \lambda^{1/4} z_\star}
\end {equation}
for fixed $\tau$ (and so fixed $z$) in the classical regime.
Using (\ref{eq:stop}) and (\ref{eq:zstar}), this may be written as
\begin {equation}
   \bigl|X^\three_n\bigr|_{\rm rms}
   \simeq
   \frac{3^{1/3} (8\pi)^{1/2}\, C(\xi_n)}{n^{1/2}\,\Gamma^2(\tfrac14)}
    \, \lambda^{-1/4} \lstop
\label {eq:Xrms}
\end {equation}
for $z \gg z_\star,\xi_n$.

Note that fixed-$\tau$ (i.e.\ fixed-$z$) slices of the string worldsheet
look different than fixed-$x^0$ slices of the string worldsheet,
which is why our depiction of the string at various times $x^0$
in fig.\ \ref{fig:boost}c were not horizontal.


\subsection {The size of the string at late times}

Parametrically, the average size (\ref{eq:Xrms})
of each stretched mode $X^\three_n$ in Poincare coordinates is just
the $\lambda^{-1/4} \lstop$ that we found in our back-of-the-envelope
argument in section \ref{sec:envelope}.
But we should now look at what happens if we sum up all the modes
to get the total average size of the string.
A convenient measure of the size scale of the string is the average rms
deviation from the center of the string,
\begin {equation}
   (\delta X^\three)_{\rm rms}
   \equiv \biggl\langle\>
            \overline{\bigl(
                X^\three - \overline{X^\three} \,
            \bigr)^2 \,}
          \>\biggr\rangle^{1/2} ,
\end {equation}
where overlines indicate averaging over the string worldsheet position
$\sigma$ and the angle brackets indicate averaging over the late-time
classical probability distribution for each mode amplitude.
This is given by
\begin {equation}
   (\delta X^\three)_{\rm rms}
   = \biggl(
        2 \sum_{n=1}^\infty \bigl\langle \bigl|X^\three_n\bigr|^2 \bigr\rangle
     \biggr)^{1/2} .
\label {eq:size0}
\end {equation}
$\bigl\langle \bigl|X^\three_n\bigr|^2 \bigr\rangle$ is just the square of
what we called $\bigl|X^\three_n\bigr|_{\rm rms}$ in (\ref{eq:Xrms}).
Combining the limiting forms (\ref{eq:Cxilimits}) with
(\ref{eq:Xrms}), and recalling from (\ref{eq:xidef}) that
$\xi_n = n^{1/3} \xi_1$,
\begin {equation}
   |X^\three_n\bigr|^2_{\rm rms} \simeq
   \left( \frac{3^{1/3} (8\pi)^{1/2}}{\Gamma^2(\tfrac14)}
          \, \lambda^{-1/4} \lstop \right)^2 \times
   \begin {cases}
      \frac{[C(0)]^2}{n} , & n \ll n_\star ; \\
      \frac{\Gamma^2(\tfrac56)}{\pi \xi_1^2 n^{5/3}} , & n \gg n_\star ,
   \end {cases}
\label {eq:Xcases}
\end {equation}
where $C(0)$ is given by (\ref{eq:C0}).
The sum in (\ref{eq:size0}) is therefore convergent
at large $n$ (more on that in a moment)
and is dominated by a logarithm coming from $n=1$ up to $n \sim n_\star$.
At leading order in inverse powers of that logarithm, 
\begin {equation}
   (\delta X^\three)_{\rm rms}
   \simeq |X^\three_1|_{\rm rms} \sqrt{2 \ln n_\star}
   \simeq
   \frac{3^{1/3} 4\sqrt\pi \, C(0)}{\Gamma^2(\tfrac14)}
          \, \lambda^{-1/4} \lstop \sqrt{\ln n_\star} .
\label {eq:result0}
\end {equation}
Using (\ref{eq:nstar}), this may be rewritten as%
\footnote{
  A comment on randomness:
  $|X_n^\three|_{\rm rms}$, determined by (\ref{eq:Xcases}), is
  the width of a Gaussian distribution for $X_n^\three$ that is centered
  on $X_n^\three = 0$.  In contrast, at leading log order,
  the $\sigma$-averaged extent
  ${\cal S} \equiv \Bigl[\,\overline{\bigl(
                X^\three - \overline{X^\three} \,
            \bigr)^2 \,}\,\Bigr]^{1/2}$
  of the late-time string is {\it always}\/ given by the
  $(\delta X^\three)_{\rm rms}$ of (\ref{eq:result}).
  That's because, when the logarithm is large,
  the calculation of the logarithm in (\ref{eq:result}) involves a sum
  over a large number $n_\star$ of modes, all contributing to
  the coefficient of the leading log, and so the
  probability distribution of ${\cal S}^2$ is narrowly peaked about
  its probabilistic average
  $(\delta X^\three)_{\rm rms}^2 \equiv \langle {\cal S}^2 \rangle$ by
  the central limit theorem.  The $n{=}1$ mode by itself is a
  substantial piece of the first correction beyond leading-log order,
  and so randomness will enter if one moves beyond a leading-log
  analysis of ${\cal S}^2$.
}
\begin {equation}
   (\delta X^\three)_{\rm rms}
   \simeq
   0.8660 \, \lambda^{-1/4} \lstop \,
    \ln^{1/2}\biggl(\frac{\lambda^{-1/6} \lmax}{\lstop}\biggr) ,
\label {eq:result}
\end {equation}
where a parametric expression for the argument of the logarithm is
adequate if we are only keeping track of the logarithmic term.

Ignoring the numerical constant in front, (\ref{eq:result})
is the parametric result (\ref{eq:ratio}) that we presented in
the introduction.
One could go on to evaluate the non-logarithmic corrections to
(\ref{eq:result}), but
(\ref{eq:result}) is good enough for our present purposes.  We are
mostly interested in the parametric size of the answer, so that
we can determine whether the extent of the string in $x^\three$
remains small compared to the stopping distance $\lstop$ in
the case of large but finite $\lambda$.

Qualitatively,
what is the origin of the logarithmic factor in (\ref{eq:result0})
and (\ref{eq:result})?
Imagine doing the same calculation of $(\delta X^\three)_{\rm rms}$
for the graviton instead of for the stretched, classical string.
In the ground state, all the modes $X^i_n$ have
an rms size proportional to $n^{-1/2}$ and so the sum
$\sum_n \langle |X^i_n|^2 \rangle$ is log divergent in the
ultraviolet.  This does not mean, however, that the energy or
momentum carried by the graviton is smeared over infinite
spatial distances:%
\footnote{
  For some discussion of the unobservability of the
  UV log divergence and
  and how the ``size'' of a low-mass string excitation should be
  interpreted, see refs.\ \cite{size}.
}
in the string's ground state,
the bosonic $n\not=0$ mode contributions to
energy and momentum are canceled by the fermionic mode contributions,
which we have ignored in our analysis.
Only the bosonic mode amplitudes, however, can become ``big'' due to
tidal stretching, and it is the modes whose amplitudes have grown big
that we consider when we make the classical approximation at late times.
Our expression (\ref{eq:Xcases}) is only valid until $n$ gets so large
that the bosonic mode
has not been significantly excited
($\xi_n \sim \bar z$, and so
$n \sim n_\star (z/z_\star)^3 \gg n_\star$ when $z \gg z_\star$).
For yet larger $n$ the
modes will be in their ground state and the canceling contribution
of fermionic modes to physical quantities of interest will come into
play.  The contribution of such
high modes is sub-leading and is simply discarded when we approximate
the string as a classical string at late times.
In the resulting treatment of the classical string, there is no
ambiguity in what the size of the string means,
as highlighted by the convergence of the sum of (\ref{eq:Xcases}).%
\footnote{
  We point out in passing that this is not an obscure issue specific
  to relativistic strings.  If one quantizes small transverse
  vibrations of an idealized violin string
  $\bigl(L = \int dz [ \tfrac12 \rho \, \dot\x_\perp^2 
              + \tfrac12 \tension \, \x_\perp^{\prime 2}] \bigr)$,
  the same $\sum_n n^{-1}$ divergence arises in
  the calculation of the mean-square displacement of
  the string.  But, if you took a snapshot in time of
  an actual vibrating violin
  string (with classically large amplitude), you would not have
  any real-world confusion about what the mean-square displacement of that
  string meant.  You would only get confused if you perversely chose to
  resolve the (idealized) violin string on such small distance scales
  that you could see the quantum uncertainty of very high modes that had not
  been excited, and for such measurements the classical description of the
  violin string would be inadequate.  If you're not interested
  in such high resolution, then the classical result for the
  average displacement is an excellent approximation and is
  physically meaningful information.
}

The upshot is that the logarithm in (\ref{eq:result0})
arises because of (i) the logarithmic UV divergence associated with
the bosonic modes in the ground state, combined with (ii) the fact that
the bosonic mode amplitudes with $n \ll n_\star$ all grow by an equal
large factor from tidal stretching (and no longer cancel against
fermionic modes in their physical consequences),
while those with $n \gg n_\star$ do not grow significantly in comparison.


\section{Discussion of the single graviton approximation}
\label {sec:gravitons}

We have followed the evolution
of a single graviton as it is stretched into a classical string loop.
We will now take a moment to discuss in more detail the premise that
we may follow a single graviton.

First, a single high-momentum string loop might split into two, but such
splitting is $1/\Nc$ suppressed and so may be ignored in the
$\Nc{=}\infty$ limit that we take in this paper.

The same argument might be given against the
possibility of two gravitons merging
if not for the fact that the localized gravitational
wavepacket describing our excitation in the bulk is classical, which
means it contains a correspondingly {\it large}\/ number of gravitons.
We need a different argument for why interactions between the gravitons
that make up the classical wavepacket may be ignored.
For this, we need to review in slightly more detail the formalism
of refs.\ \cite{adsjet,adsjet2} for creating the excitation in
the first place.  In that formalism, the initial photon or W boson
or whatever of fig.\ \ref{fig:Wdecay} is replaced by a localized,
external classical field.
Specifically, we add a source term to the Lagrangian,
\begin {equation}
   {\cal L} \to
   {\cal L} + {\cal N} \, O(x) \, e^{i\bar k\cdot x} \, \Lambda_L(x) ,
\label {eq:source}
\end {equation}
where ${\cal N}$ is an arbitrarily small source amplitude,
$O(x)$ is a source operator
(corresponding to the vertex in fig.\ \ref{fig:Wdecay}),
\begin {equation}
   \bar k^\mu \simeq (E,0,0,E)
\end {equation}
is the large 4-momentum of the desired excitation, and
$\Lambda_L(x)$ is a slowly varying envelope function that localizes
the source near the origin in both $x^\three$ and time.
The amplitude ${\cal N}$ is chosen to be small
so that we can treat the external
source as a small perturbation to the strongly-interacting gauge theory,
so that the source will never create more than one
jet with energy $E$ at a time.
On an event-by-event basis, the source will usually do
nothing at all, but on rare occasions ($\propto |{\cal N}|^2$), it
will create an excitation with energy $\simeq E$.
For small enough ${\cal N}$, it will essentially never create excitations
with energies $2E$, $3E$, etc.\ \cite{adsjet2}.
But taking ${\cal N}$ small also means that the bulk excitation
created by the source on the boundary can be treated in linear response:
the self-interactions of the bulk excitation with itself are ignorable.
For this reason, we may ignore interactions between the gravitons
(or other type of particles) that make up the bulk excitation.

Some readers may wonder how we can focus on gravitons, which are
quantum mechanical objects, when the
dual theory for $\Nc{=}\infty$ is supposed to be a classical theory.
As an analogy, consider a classical electromagnetic wave with
polarization $({\bm e}_x+{\bm e}_y)/\sqrt2$.
If we choose to, we may think of this classical wave as a coherent
superposition of photons which are in the quantum state
$\bigl(|x\rangle + |y\rangle\bigr)/\sqrt2$, where $|x\rangle$ and
$|y\rangle$ are states corresponding to polarization in the $x$ and
$y$ directions.  Now measure the polarization by putting the wave
through a linear polarizer oriented in the $x$ direction.
As we all know, we may view this classically as picking out the
${\bm e}_x/\sqrt2$ component of the wave, or alternatively
we may view it in
terms of photons as saying that each photon has probability
$1/2$ of being $x$-polarized.  A discussion of classical physics
in one description is equivalent to a quantum-mechanical
discussion of probabilities for the behavior of individual quanta
in the other description.

The same reasoning applies to our description of the classical bulk
wavepacket in terms of individual gravitons.
An important conceptual lesson from this is that our wavepacket
is not a single graviton that evolves
into a single classical string loop
with probability distribution given by a Gaussian with size
(\ref{eq:result}) for each mode $n$.  Instead, the wavepacket is a
large number of gravitons that independently evolve into a large
number of classical string loops that independently have that
probability distribution.
Following the analogy with photon polarizations further, one could presumably
replace the quantum description involving probability amplitudes
for the degrees of freedom of individual string loops by
a classical description in terms of a classical string field theory,
promoting the Schr\"odinger wave functional $\psi[{X_n^i}]$ for
a single string to a classical wave functional
$\psi[{X_n^i}]$ that could be used to describe all the physics
discussed in this paper.  We have not pursued this latter approach
because
we thought that the description in terms of gravitons was simpler,
more intuitive,
and more directly related to the previous literature on tidal
excitation.


\section {Checking the Penrose limit}
\label {sec:PenroseCheck}

Now that we have our final answer (\ref{eq:result})
for the size of the classical
string that is produced by stretching,
we should go back and verify that the string is not so big, or
so far away from the reference geodesic, that
the Penrose limit taken in section \ref{sec:Penrose} breaks down.
We need to check that the $dv\>d\Delta x^\three$ and $(dv)^2$ terms
in the AdS$_5$-Schwarzschild metric (\ref{eq:metricuv}),
which were dropped in the Penrose limit, are parametrically
small compared to the $du\>dv$ term,
\begin {equation}
   \frac{f\Rads^2}{z^2} \frac{|\q|}{\omega} \> |dv \> d\Delta x^\three|
   \quad \mbox{and} \quad
   \frac{f\Rads^2}{z^2} \> (dv)^2
   \quad \ll \quad
   |du\>dv| ,
\end {equation}
for the string motions that we have found.  Dividing both sides by
$|du\>dv|$ and using $\omega \simeq |\q|$, we may rewrite these conditions
as
\begin {equation}
   \frac{f\Rads^2}{z^2} \left|\frac{d\Delta x^\three}{du}\right|
   \quad \mbox{and} \quad
   \frac{f\Rads^2}{z^2} \left|\frac{dv}{du} \right|
   \quad \ll \quad
   1 .
\label {eq:PenroseConstraints}
\end {equation}
We check these conditions on the string motion in appendix
\ref{app:PenroseCheck}, where we find that the condition on
$|d\Delta x^\three/du|$ is the strongest and requires
\begin {equation}
   \lambda^{-1/4} \sqrt{\ln n_\star} \ll 1
\label {eq:PenroseFinal0}
\end {equation}
in order for our earlier analysis to be valid.  Using our result
(\ref{eq:result0}), this condition may be written as
\begin {equation}
   (\delta X^\three)_{\rm rms} \ll \lstop .
\label {eq:PenroseFinal}
\end {equation}
That is, the Penrose limit only breaks down if one considers the
extreme case (to be discussed in a moment) where
the string becomes as large as the stopping distance itself.


\section {Conclusion}
\label {sec:conclusion}

In our scheme for creating ``jets,'' we have seen different behaviors
in the dual theory depending on the virtuality (and so the stopping
distance) of the jet.
For $\lstop \gg \lambda^{-1/6}\lmax$, the gravitons (or other massless
string modes) composing the excitation in the gravity description remain
gravitons until after the excitation has stopped moving in the
$x^\three$ direction, and there is no
difficulty in using the supergravity approximation for the calculation.
For $\lstop \ll \lambda^{-1/6}\lmax$, each graviton is instead
stretched into a classical string loop.  However, provided that
\begin {equation}
   \lambda^{-1/4} \, \ln^{1/2}\biggl(\frac{\lambda^{-1/6} \lmax}{\lstop}\biggr)
   \ll 1 ,
\end {equation}
the size of that string remains small compared to the stopping distance
$\lstop$.  The string remains close to its reference geodesic, and so
corrections to the $\lambda{=}\infty$ description of the
jet stopping are parametrically small (if one only attempts to resolve
details on size scales large compared to the size of the string).
However, if instead
\begin {equation}
   \lambda^{-1/4} \, \ln^{1/2}\biggl(\frac{\lambda^{-1/6} \lmax}{\lstop}\biggr)
   \gtrsim 1 ,
\label {eq:biglog}
\end {equation}
then the string loop will stretch out to be parametrically as large
as the stopping distance itself.  Our quantum analysis of the string
breaks down in this case (because of the failure of the Penrose limit),
but we can see what happens qualitatively by tracking what happens
to the $\lambda^{-1/4} \log^{1/2} \ll 1$ results as we increase the logarithm
towards $\lambda^{-1/4} \log^{1/2} \sim 1$.

\begin {figure}
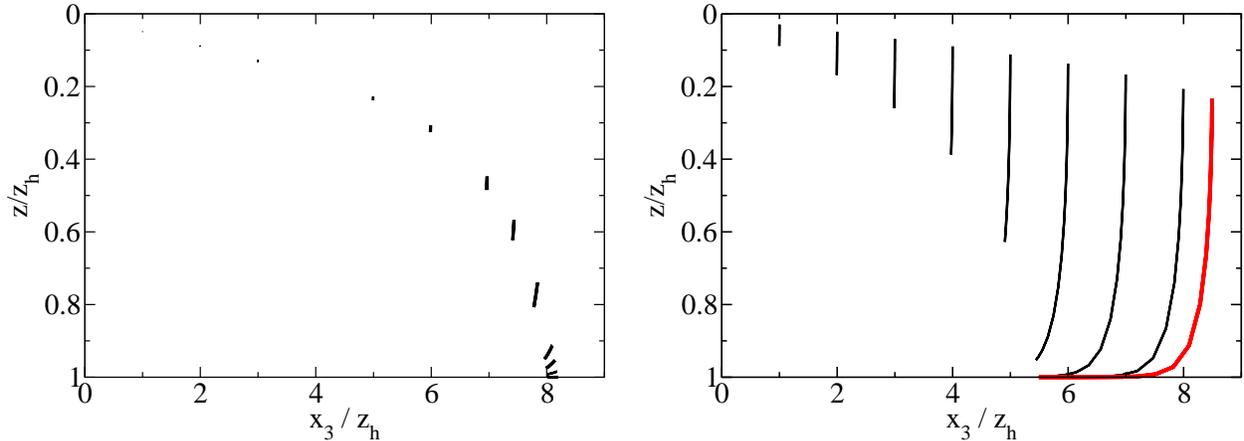

\begin {center}
  \includegraphics[scale=0.32]{string1.eps}
  \hspace{0.1in}
  \includegraphics[scale=0.32]{string2.eps}
  \caption{
     \label{fig:string}
     Examples of numerical solutions of the evolution of a
     falling classical string
     loop that starts near the boundary with proper size
     (a) $\Sigma \ll \Rads$
     and (b) $\Sigma \sim \Rads$.
     These are snapshots of the string at fixed $x^\zero$.
     See appendix \ref{app:string} for details of
     the initial condition.
  }
\end {center}
\end {figure}

\begin {figure}
\begin {center}
  \includegraphics[scale=0.5]{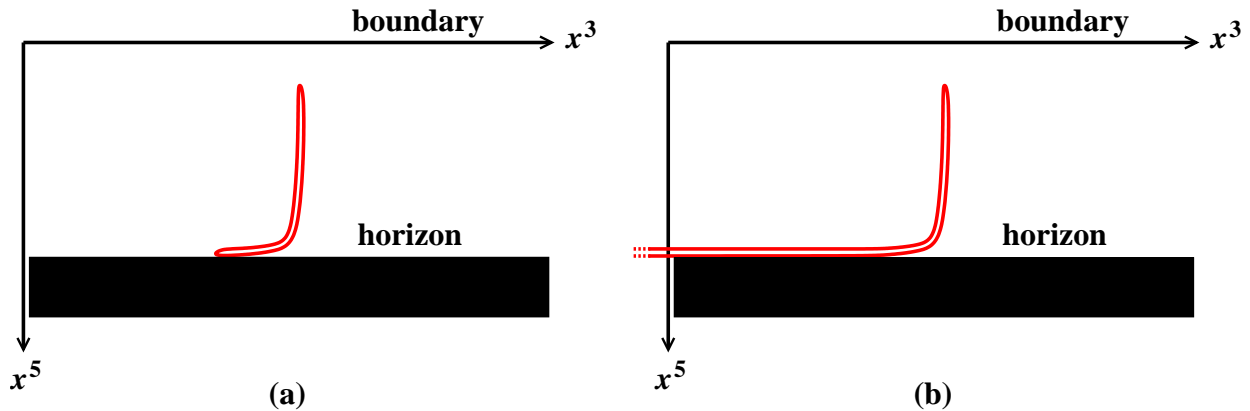}
  \caption{
     \label{fig:fold}
     Schematic pictures of classical folded strings.
     (a) A closed folded string produced by extreme tidal stretching
     of a graviton in our method of generating ``jets'' in the
     case of $\lambda^{-1/4} \log^{1/2} \gtrsim 1$ (\ref{eq:biglog}),
     and (b) the infinite, folded open string considered by
     Gubser et al.\ \cite{GubserGluon}.  In the latter case,
     the trailing string continues to get closer and closer to the
     horizon as $x^\three \to -\infty$.
  }
\end {center}
\end {figure}

In particular, a nice way to visualize what happens is to follow the
classical evolution of a closed string that initially starts with
a proper size $\Sigma$ of order $\sqrt{\alpha' \ln n_\star}$, which is
roughly the
initial rms size from the modes $n \lesssim n_\star$ which become
classically excited.  Increasing the logarithm towards
$\lambda^{-1/4} \sqrt{\ln n\star} \sim 1$ is equivalent to increasing
$\Sigma$ towards $\sim \Rads$.  Fig.\ \ref{fig:string} compares
examples of such evolution for the cases (a)
$\lambda^{-1/4} \log^{1/2} \ll 1$ and (b) $\lambda^{-1/4} \log^{1/2} \sim 1$.
(More details of exactly how we initialized our classical string may be
found in appendix \ref{app:string}.)
The interesting feature of fig.\ \ref{fig:string}b is that, at
late times, the string looks like the original picture advocated by Gubser
et al.\/ \cite{GubserGluon} of gluon jets as dual to the evolution of
a trailing, folded classical string falling in AdS$_5$-Schwarzschild.
Our string is a folded closed string, as depicted in the cartoon of
fig.\ \ref{fig:fold}a, whereas the one studied by
Gubser et al.\ was a folded infinite open string,
as depicted
by fig.\ \ref{fig:fold}b.%
\footnote{
  More precisely, Gubser et al.\ first considered a folded open string
  that stretched out from beyond the horizon, as in fig.\ 1 of
  ref.\ \cite{GubserGluon}.  But in actual calculations, they focused
  on the trailing infinite folded string, as in fig.\ 2 of that
  reference.
}
However, the left end of the string in
these figures,
which is very close to the horizon, does not play a significant role
in the effect on the boundary theory, and so the physics of these
two situations is approximately the same.

Historically, the original motivation of our own method
for posing ``jet'' stopping problems \cite{adsjet},
outlined in the introduction to this paper and
motivated by fig.\ \ref{fig:Wdecay}, was to give
a precise field theory problem in
${\cal N}{=}4$ SYM that could
be solved, beginning to end, using gauge-gravity duality.
It has not previously been know how to precisely set up a problem
in ${\cal N}{=}4$ SYM that corresponds to earlier studies
of jets \cite{GubserGluon,HIM,CheslerQuark}
that made use of classical strings in the gravity
description.  It is interesting to now make contact between our
approach and Gubser et al.'s classical string approach, in the particular limit
(\ref{eq:biglog}), which can be roughly rewritten as
\begin {equation}
   T^{-1} \ll \lstop
   \lesssim
   T^{-4/3} \left( \frac{E}{\sqrt\lambda} \right)^{1/3} e^{-O(\lambda^{1/2})} ,
\label {eq:overlap}
\end {equation}
where the $T^{-1} \ll \lstop$ is thrown in to emphasize that we've always been
assuming $-q^2 \ll E^2$ and so $T^{-1} \ll \lstop$ throughout,
and where $O$ means ``of order.''
Alternatively, in terms of the virtuality $-q^2$ of the source
of our ``jet,''
(\ref{eq:overlap}) is
\begin {equation} 
   E^2 \gg -q^2 \gtrsim
   T^{4/3} E^{2/3} \lambda^{2/3} e^{+O(\lambda^{1/2})}
   .
\end {equation}
This window of stopping lengths only appears once the jet energy is large
enough that
\begin {equation}
   E \gg T \sqrt{\lambda} \> e^{+O(\lambda^{1/2})} .
\end {equation}

Even though there is a region of overlap (\ref{eq:overlap})
of our results with strings that look similar to those of
Gubser et al., there are still important differences
once we get out of this range.
Gubser et al.\ found a {\it maximum}\/
stopping distance of order $T^{-4/3} (E/\sqrt{\lambda})^{1/3}$, as do
other methods that also model excitations with semi-infinite classical strings
in the gravity dual \cite{CheslerQuark}.  In contrast, the
types of excitations that we create, through processes like
fig.\ \ref{fig:Wdecay}, have a parametrically larger
maximum stopping distance of order $\lmax \sim T^{-4/3} E^{1/3}$.

In this paper, we have studied the case of large but finite $\lambda$,
but we have kept $\Nc{=}\infty$.  It is important to
mention, following Gubser et al.\ \cite{GubserGluon},
that the case of finite $\Nc$ may be qualitatively different.
For finite $\Nc$, the folded classical string may break,
creating many string loops.  It would be interesting to understand
whether or not such breaking would impact the
description of jet stopping in ${\cal N}{=}4$ SYM.

Our basic argument in this paper has been that the breakdown of the
$1/\sqrt{\lambda}$ expansion seen in the earlier supergravity
calculation of ref.\ \cite{R4} is explained by the tidal stretching of
gravitons into classical string loops.  Logically, however, to show
that tidal stretching was the only source of difficulty, we should now
do a systematic analysis of sub-leading corrections to the string
evolution that we have presented here and show that those corrections
are indeed parametrically small.
We expect that they will be, but we leave that for someone to study
another day.

Finally, we mention that an alternative to our method
for creating ``jets'' is to produce gluon jets as synchrotron
radiation from heavy quarks that are forced into circular motion.
The corresponding jet stopping problem has been investigated for
strongly-coupled ${\cal N}{=}4$ SYM by Chesler,
Ho, and Rajagopal \cite{CHR}.
We suspect that problems similar to those treated here could be set up
in that context as well, with appropriate choices of the heavy quark
velocity and synchrotron radius for a given temperature $T$.
It would be interesting to investigate.


\begin{acknowledgments}

This work was supported, in part, by the U.S. Department
of Energy under Grant No.~DE-SC0007984.

\end{acknowledgments}


\appendix

\section{Large \boldmath$\xi$ behavior of \boldmath$C(\xi)$}
\label{app:Cxibig}

For large $\xi$, the $\bar z^6$ term in the differential equation
(\ref{eq:chixi}) for $\xi$ can be ignored until $\bar z \gg 1$.
At that point, however, we may use the simple large-$\bar z$
result (\ref{eq:zbarlate}) for $\bar z$.  Substituting this into
(\ref{eq:chixi}) gives
\begin {equation}
   \frac{d^2\chi}{d\bar\tau^2}
   = -4\left[\xi^6 - \frac{1}{(-3\bar\tau)^2}\right] \chi ,
\end {equation}
whose solution is
\begin {equation}
   \chi =
   (-\pi \xi^3 \bar\tau)^{1/2} H_{5/6}^{(2)}\bigl(-2 \xi^3 \bar\tau \bigr) .
\end {equation}
The late-time behavior $\tau\to 0$ is
\begin {equation}
   |\chi| \simeq
   \frac{\Gamma(\tfrac56)}{\pi^{1/2} \xi (-\bar\tau)^{1/3}}
   \,.
\end {equation}


\section{Checking the Penrose limit: details}
\label{app:PenroseCheck}

In this appendix, we will check whether the evolution of our strings
satisfy the conditions (\ref{eq:PenroseConstraints}) required for
taking the Penrose limit.
It is illuminating to do this in two different ways.
First, we will make a rough argument based on following divergent
geodesics, similar to the style of argument that we used in
section \ref{sec:envelope}.
Then we will give alternative (but physically equivalent)
arguments based on string-based results and formulas.


\subsection{Tracking diverging geodesics}

As in section (\ref{sec:envelope}), let us characterize the string by
following null geodesics that roughly trace different bits of string and
which deviate slightly from our reference geodesic.  As we'll discuss
again later, this approximation amounts to ignoring the tension in
the string.  From the null geodesic formula
(\ref{eq:geodesic}) and the metric (\ref{eq:metric}), the
$x^\three$ coordinate for such geodesics is given by
\begin {equation}
   \frac{dx^\three}{dz} = \frac{\hat q_\three}{\sqrt{1-f\hat\q^2}} ,
\label {eq:dDx3a}
\end {equation}
where
\begin {equation}
   \hat q_\mu \equiv \frac{q_\mu}{\omega} = (-1,\hat\q) .
\end {equation}
Remembering that $\Delta x^\mu \equiv x^\mu - \bar x^\mu(z)$
is the deviation relative to the reference geodesic, we have
\begin {equation}
   \frac{d\Delta x^\three}{dz} =
   \frac{\hat q_\three}{\sqrt{1-f\hat\q^2}}
   - \frac{\bar{\hat q}_\three}{\sqrt{1-f\bar{\hat\q}^2}}
   .
\end {equation}
Expand to first order in $\Delta\hat q_3 \equiv \hat q_3 - \bar{\hat q}_3$:
\begin {equation}
   \frac{d\Delta x^\three}{dz} \simeq
   \frac{\Delta\hat q_\three}{(1-f\bar{\hat\q}^2)^{3/2}}
   .
\label {eq:dv1}
\end {equation}
Then using (\ref{eq:partialu}) (and defining $u$ with respect to the
reference geodesic $\bar x$),
\begin {equation}
   \frac{f \Rads^2}{z^2} \, \frac{d\Delta x^\three}{du} \simeq
   \frac{f \, \Delta\hat q_\three}{1-f\hat\q^2}
   .
\label {eq:combo1}
\end {equation}
Since $1-f\hat\q^2 \simeq (z_\star^4+z^4)/\zh^4$, the
combination (\ref{eq:combo1}) is largest for $z \lesssim z_\star$,
and the $d\Delta x^\three/du$ condition in
(\ref{eq:PenroseConstraints}) requires
\begin {equation}
   \Delta \hat q_\three \ll \frac{z_\star^4}{\zh^4}
\label {eq:cond1a}
\end {equation}
for the Penrose limit.
Use (\ref{eq:stop}) to relate this to the stopping distance:
\begin {equation}
   \lstop  \sim \frac{\zh^2}{z_\star}
   \sim \zh \left( \frac{E^2}{-q^2} \right)^{1/4}
   \sim \frac{\zh}{(1-\hat q_\three)^{1/4}} ,
\end {equation}
so that
\begin {equation}
   \Delta\lstop
   \sim \frac{\zh\,\Delta\hat q_\three}{(1-\hat q_\three)^{5/4}}
   \sim \frac{\Delta\hat q_\three \, \lstop}{1-\hat q_\three}
   \sim \Delta\hat q_\three \, \lstop \, \frac{\zh^4}{z_\star^4} .
\label {eq:cond1b}
\end {equation}
Combining (\ref{eq:cond1a}) and (\ref{eq:cond1b}) gives the condition
\begin {equation}
   \Delta\lstop \ll \lstop
\label {eq:condition}
\end {equation}
quoted in (\ref{eq:PenroseFinal}).

Now turn to the condition on $dv/du$ in (\ref{eq:PenroseConstraints}).
The definition (\ref{eq:v}) of $v$ gives
\begin {equation}
    dv = \bar{\hat q}_\mu \, d(\Delta x^\mu)
    = - d(\Delta x^\zero) + \bar{\hat q}_\three \, d(\Delta x^\three),
\label {eq:dv2}
\end {equation}
and so we need a formula for $d(\Delta x^\zero)$.
The analog of (\ref{eq:dDx3a}) is
\begin {equation}
   \frac{dx^\zero}{dz} = \frac{- f^{-1} \hat q_\zero}{\sqrt{1-f\hat\q^2}} ,
\end {equation}
with expansion
\begin {equation}
   \frac{d\Delta x^\zero}{dz} \simeq
   \frac{\hat q_\three \, \Delta\hat q_\three}{(1-f\bar{\hat q}^2)^{3/2}}
   .
\label {eq:dv3}
\end {equation}
Combining (\ref{eq:dv1}), (\ref{eq:dv2}), and (\ref{eq:dv3}),
gives $dv/du \simeq 0$.  We therefore have to go back and make
our expansions to second-order in $\Delta\hat\q$.  The result is
\begin {equation}
  \frac{dv}{dz} \simeq
  - \frac{(\Delta\hat\q_\perp)^2}{2(1-f\bar{\hat q}_3^2)^{1/2}}
  - \frac{(\Delta \hat q_\three)^2}{2(1-f \bar{\hat q}_3^2)^{3/2}}
  \,,
\end {equation}
and so
\begin {equation}
  \frac{f\Rads^2}{z^2} \left|\frac{dv}{du}\right| \simeq
  \frac{f(\Delta\hat\q_\perp)^2}{2}
  + \frac{f(\Delta \hat q_\three)^2}{2(1-f \bar{\hat q}_3^2)}
  \,.
\end {equation}
The corresponding condition on $dv/du$ in (\ref{eq:PenroseConstraints})
is then
\begin {equation}
  f(\Delta\hat\q_\perp)^2
  \quad \mbox{and} \quad
  \frac{f(\Delta \hat q_\three)^2}{(1-f \bar{\hat q}_3^2)}
  \quad \ll \quad
  1 .
\label {eq:dvdugeo}
\end {equation}
The first condition is strongest for $z \ll \zh$ and the second for
$z \lesssim z_\star$, giving
\begin {equation}
  |\Delta\hat\q_\perp|
  \quad \mbox{and} \quad
  |\Delta \hat q_\three| \, \frac{\zh^2}{z_\star^2}
  \quad \ll \quad
  1 .
\label {eq:cond2}
\end {equation}
Using (\ref{eq:cond1b}), the condition involving $\Delta\hat q_\three$
becomes
\begin {equation}
  \Delta\lstop \ll \lstop \frac{\zh^2}{z_\star^2} \,.
\label {eq:condition9}
\end {equation}
Since $z_\star \ll \zh$, this is weaker than the previous condition
(\ref{eq:condition}).

Now turn to the other condition,
$|\Delta\q_\perp| \ll 1$ in (\ref{eq:cond2}).
To estimate $|\Delta\q_\perp|$, return to the
arguments of section \ref{sec:envelope}, but now, in the
rest frame, include an initial
proper displacement of the two geodesics
in $\x^\perp$ of the same parametric size as the initial
proper displacement in $x^\three$.  Following through the argument, one finds
\begin {equation}
   x^I \simeq
   \Bigl(\gamma(1 + \beta\beta_+),{\bm\beta}_\perp,
         \gamma(\beta + \beta_+),1\Bigr) \, z
\end {equation}
with $\beta_\perp \sim \beta_+$.
Then
\begin {equation}
   \Delta \hat q_\three
   = \Delta \frac{q_\three}{q_\zero}
   = \Delta \frac{dx^\three/dz}{dx^\zero/dz}
   \simeq \Delta \frac{\gamma(\beta + \beta_+)}{\gamma(1+\beta\beta_+)}
   \simeq \frac{\beta_+}{\gamma^2}
\end {equation}
and
\begin {equation}
   \Delta \hat\q_\perp
   = \Delta \frac{\q_\perp}{q_\zero}
   = \Delta \frac{d\x^\perp/dz}{dx^\zero/dz}
   \simeq \Delta \frac{{\bm\beta}_\perp}{\gamma(1+\beta\beta_+)}
   \simeq \frac{{\bm\beta}_\perp}{\gamma} ,
\end {equation}
so that
\begin {equation}
   \frac{|\Delta\hat\q_\perp|}{|\Delta\hat q_\three|}
   \sim \gamma \sim \sqrt{\frac{E^2}{-q^2}} \sim \frac{\zh^2}{z_\star^2}
   \,.
\end {equation}
So, using (\ref{eq:cond1b}),
\begin {equation}
  |\Delta\hat\q_\perp| \sim |\Delta\hat q_\three| \, \frac{z^2}{z_\star^2}
  \sim \frac{\Delta\lstop}{\lstop} \, \frac{z_\star^2}{\zh^2} \,.
\label {eq:qperpidentify}
\end {equation}
The condition $|\Delta\hat\q_\perp| \ll 1$ is therefore the same
as the previous condition (\ref{eq:condition9}) and so is
also weaker than (\ref{eq:condition}).


\subsection{String-based formulas}
\label{app:PenroseCheckString}

Another way to get to the same result is to start from formulas based
on our string analysis in section \ref{sec:quantize}.  We will also need
the constraint equation for $\hat V$ generated by (\ref{eq:action})
in $\tau=U$ gauge,
which is
\begin {equation}
  \partial_\tau \hat V = 
  -\tfrac12 \sum_i \left[
     (\partial_\tau \Delta\hat X^i)^2
     + \frac{1}{(\alpha' p^u)^2} (\partial_\sigma \Delta\hat X^i)^2
     + {\cal G}_i \, (\Delta\hat X^i)^2
   \right] .
\end {equation}
Using the definition of (\ref{eq:vhat}) of $\hat v$, the corresponding
constraint for $V$ is
\begin {equation}
  \partial_\tau V = 
  -\tfrac12 \sum_i \left[
    \kappa_i \Bigl(\partial_\tau \frac{\Delta\hat X^i}{\sqrt{\kappa_i}}\Bigr)^2
     + \frac{1}{(\alpha' p^u)^2} (\partial_\sigma \Delta\hat X^i)^2
   \right] ,
\label {eq:dvstring}
\end {equation}
where $\kappa_i = \kappa_i(\tau)$ is the notation introduced in eq.\
(\ref{eq:metricyhat}) with
\begin {equation}
  \kappa_1 = \kappa_2 = \frac{\Rads^2}{z^2} \,,
  \qquad
  \kappa_3 = \frac{\Rads^2}{z^2} \frac{(\omega^2-f|\q|^2)}{\omega^2}
           \simeq \frac{\Rads^2}{z^2} \, \frac{(z_\star^4+z^4)}{\zh^4} .
\label {eq:kappa}
\end {equation}
The relationship (\ref{eq:yhat}) between $\Delta X^i$ and $\Delta\hat X^i$
is
\begin {equation}
   \Delta\hat X^i \equiv \sqrt{\kappa_i} \, \Delta X^i .
\label {eq:Xhat}
\end {equation}
Alternatively, (\ref{eq:dvstring}) is just the constraint equation
$T_{\tau\tau}=0$ that one would derive in Rosen coordinates
(\ref{eq:metricpp0}).

Here and throughout, we will take results derived in the Penrose limit
but then use them to check, after the fact, whether the Penrose limit was
valid.
We will focus on the evolution of the string modes
after they become unstable ($z \gg z_n$).
The string
equation of motion from the Lagrangian (\ref{eq:Lin}) is
\begin {equation}
  \partial_\tau^2 \Delta\hat X_n^i =
  - \omega_{i,n}^2(\tau) \, \Delta\hat X_n^i ,
\end {equation}
where $\omega_{i,n}^2(\tau) = n^2(\alpha'p^u)^{-2} - {\cal G}_i(\tau)$.
At late times $z \gg z_n$, the tidal terms ${\cal G}_i$ dominate over the
string tension terms $n^2(\alpha' p^u)^{-2}$, and the equation of motion
becomes
\begin {equation}
  \partial_\tau^2 \Delta\hat X_n^i \simeq
  {\cal G}_i \, \Delta\hat X_n^i ,
\end {equation}
which it is convenient to rewrite as
\begin {equation}
  \partial_\tau^2 \Delta\hat X_n^i \simeq
  \frac{\partial_\tau^2 \sqrt{\kappa_i}}{\sqrt{\kappa_i}} \, \Delta\hat X_n^i .
\label {eq:EOMtensionless}
\end {equation}
Two independent solutions for $\Delta\hat X$ in the tensionless approximation
(\ref{eq:EOMtensionless}) are
\begin {equation}
  A^i(\tau) \propto \sqrt{\kappa_i(\tau)} ,
  \qquad
  B^i(\tau) \propto
  \sqrt{\kappa_i(\tau)} \int_{-\infty}^\tau \frac{d\tau'}{\kappa_i(\tau')} \,.
\label {eq:AB}
\end {equation}
We will take $A^i_n$ and $B^i_n$ to be normalized to be $\sim 1$ at
$z \sim z_n$.  The late-time ($z \gg z_n)$ string solutions will then be
given by some probabilistic superposition
\begin {equation}
   \Delta\hat X^i_n(\tau) = a^i_n \, A^i_n(\tau) + b^i_n \, B^i_n(\tau)
\end {equation}
where $a^i_n$ and $b^i_n$ are of order the size scale $\sqrt{\alpha'/n}$
of the proper size of the ground state of the $n$-th mode, which
characterizes the string for $z \lesssim z_n$.

From (\ref{eq:partialu}) or (\ref{eq:zbar}), the relationship
between $\tau$ and $z$ is
\begin {equation}
   -\tau = \zh^2 \Rads^2 \int_z^\infty \frac{dz}{z^2(z_\star^4 + z^4)^{1/2}}
   \sim
   \zh^2\Rads^2
   \begin {cases}
     \frac{1}{z_\star^2 z}\,, & z \ll z_\star ; \\[4pt]
     \frac{1}{z^3}\,,        & z \gg z_\star .
   \end {cases}
\end {equation}
Given the expressions (\ref{eq:kappa}) for $\kappa_i$,
the resulting behavior of the solution (\ref{eq:AB}) is then
\begin {align}
   A_n^\one = A_n^\two &\sim \frac{z_n}{z} \,
\\
   B_n^\one = B_n^\two &\sim 
   \begin {cases}
      1,                     & z \ll z_\star ; \\[4pt]
      \frac{z_\star}{z} \,,   & z \gg z_\star .
   \end {cases}
\\
   A_n^\three &\sim
   \begin {cases}
      \frac{z_n}{z},            & z \ll z_\star ; \\[4pt]
      \frac{z_n z}{z_\star^2}\,,  & z \gg z_\star .
   \end {cases}
\\
   B_n^\three &\sim
   \begin {cases}
      1,                       & z \ll z_\star ; \\[4pt]
      \frac{z}{z_\star}\,,      & z \gg z_\star .
   \end {cases}
\end {align}
Note that $B_n^i$ dominates over $A_n^i$ for $z \gg z_n$.
For $z \gg z_\star$, we have $B_n^\three \propto A_n^\three \propto z$,
and this behavior corresponds to the dominant $(-\bar\tau)^{-1/3}$ late-time
behavior discussed in section \ref{sec:late}.
At late times, $B^\three$ differs from a multiple of $A^\three$ by corrections
of absolute size of order $z_\star^4/z^4$,
which is the sub-leading $(-\bar\tau)^{4/3}$
late-time solution discussed in the main text just before (\ref{eq:chilate}).
A simple way to see this is to rewrite the definition of $B^i$
in (\ref{eq:AB}) as
\begin {equation}
  B^i(\tau) \propto
  \sqrt{\kappa_i(\tau)} \int_{-\infty}^0 \frac{d\tau'}{\kappa_i(\tau')}
  - \sqrt{\kappa_i(\tau)} \int_{\tau}^0 \frac{d\tau'}{\kappa_i(\tau')} \,,
\end {equation}
where the first term is proportional to $A^i(\tau)$ and the second is
(at late times) the sub-leading solution.
Similarly, at late times ($z \gg z_\star$), $B^\one$ differs from a
multiple of $A^\one$ by sub-leading corrections of order
$z_\star^2/z^2$.


\subsubsection{Condition on $d(\Delta x^\three)/du$}

Now let us investigate the Penrose limit condition
(\ref{eq:PenroseConstraints})
on $d(\Delta x^\three)/du$, which can be rewritten as
\begin {equation}
   \frac{f\Rads^2}{z^2}
   \left| \partial_\tau
          \Bigl(\frac{\Delta\hat X^\three}{\sqrt{\kappa_3}}\Bigr) \right|
   \ll 1 .
\label {eq:Cond1}
\end {equation}
The $A \propto \sqrt{\kappa}$ solutions do not give any contribution to
the $\tau$ derivative above, and so
\begin {equation}
   \partial_\tau
          \Bigl(\frac{\Delta\hat X_n^\three}{\sqrt{\kappa_3}}\Bigr)
   = b_n^\three \partial_\tau
          \Bigl(\frac{B_n^\three}{\sqrt{\kappa_3}}\Bigr)
   \sim \sqrt{\frac{\alpha'}{n}} \, \frac{z_\star^4/\Rads\zh^4}{\kappa_3}
   = \frac{\lambda^{-1/4}}{\sqrt{n}} \,
     \frac{z^2 z_\star^4}{\Rads^2 (z_\star^4+z^4)} \,.
\label {eq:dX3}
\end {equation}
The condition (\ref{eq:Cond1}) has to be satisfied for every point
$\sigma$ on the string.  But we can get a sufficient parametric condition
by taking the rms average over the modes $n \lesssim n_\star$ which
are excited:
\begin {equation}
   \frac{f\Rads^2}{z^2}
   \left| \partial_\tau
          \Bigl(\frac{\Delta\hat X^\three}{\sqrt{\kappa_3}}\Bigr)
   \right|_{\rm rms}
   \ll 1 ,
\label {eq:cond1}
\end {equation}
which then gives
\begin {equation}
  \lambda^{-1/4} \sqrt{\ln n_\star} \,
  \frac{f z_\star^4}{(z_\star^4+z^4)}
  \ll 1 .
\end {equation}
This condition is strongest for $z \lesssim z_\star$ and then gives
the condition (\ref{eq:PenroseFinal0}) quoted in the main text.

To relate this to the earlier geodesic analysis,
recall that $\Delta\hat X^i/\sqrt{\kappa_i} = \Delta X^i$ and
so note that (\ref{eq:dX3}) gave
\begin {equation}
   \frac{f\Rads^2}{z^2}
   \left| \partial_\tau \Delta X^\three \right|_{\rm rms}
   \sim 
   \lambda^{-1/4} \sqrt{\ln n_\star} \,
   \frac{f z_\star^4}{(z_\star^4+z^4)} .
\end {equation}
Comparing
to (\ref{eq:combo1}) then identifies
$\Delta \hat q_\three$ as
\begin {equation}
  \Delta \hat q_\three
  \sim \frac{z_\star^4}{\zh^4} \, \lambda^{-1/4} \sqrt{\ln n_\star}
  \sim \frac{z_\star^4}{\zh^4} \, \frac{\Delta\lstop}{\lstop} \,,
\label {eq:dq3identify}
\end {equation}
consistent with (\ref{eq:cond1b}).


\subsubsection{Condition on $dv/du$}

Now turn to the condition on $dv/du$.  The first term in
the constraint formula (\ref{eq:dvstring}), which is what remains if
one ignores the tension terms in that equation, simply reproduce
the conditions (\ref{eq:dvdugeo}) found in the earlier analysis
based on diverging geodesics.  To see this, consider the $i=3$
contribution of the first term in (\ref{eq:dvstring})
to the $dv/du$ constraint in (\ref{eq:PenroseConstraints}):
\begin {equation}
   \frac{f \Rads^2}{z^2} \, \kappa_3
   \Bigl(\partial_\tau \frac{\Delta\hat X^\three}{\sqrt{\kappa_3}}\Bigr)^2
   \ll 1 .
\label {eq:Cond2}
\end {equation}
Using (\ref{eq:kappa}) and (\ref{eq:dX3}), the contribution from
mode $n$ to the left-hand side of (\ref{eq:Cond2}) is of order
\begin {equation}
   \frac{\lambda^{-1/2}}{n} \,
   \frac{f z_\star^8}{\zh^4(z_\star^4+z^4)}
   \,.
\end {equation}
Summing over the excited modes gives
\begin {equation}
   \lambda^{-1/2} \ln(n_\star) \, \frac{f z_\star^8}{\zh^4(z_\star^4+z^4)}
   \ll 1 .
\label {eq:condition999}
\end {equation}
This is the same as the second condition in (\ref{eq:dvdugeo}),
with the identification (\ref{eq:dq3identify}) noted before.
Similarly, using (\ref{eq:kappa}) and (\ref{eq:AB}),
\begin {equation}
   \frac{f \Rads^2}{z^2} \, \kappa_\perp
   \Bigl(\partial_\tau \frac{\Delta\hat X^\perp}{\sqrt{\kappa_\perp}}\Bigr)^2
   \ll 1
\end {equation}
generates
\begin {equation}
   \lambda^{-1/2} \ln(n_\star) \, \frac{f z_\star^4}{\zh^4} \ll 1 ,
\end {equation}
which is the same as the first condition in (\ref{eq:dvdugeo}),
using the identifications of (\ref{eq:qperpidentify})
and (\ref{eq:dq3identify}).

Finally, there are the tension terms in
the constraint formula (\ref{eq:dvstring}) for $dv/du$.  One may
check that these also do not generate any constraints stronger than
(\ref{eq:condition}).


\section{Simulation details for fig.\ \ref{fig:string}}
\label{app:string}

\def\MT{\hat{\cal X}^\zero}
\def\MZ{\hat{\cal Z}}
\def\stretch{\Gamma}

Before we explain the initial conditions used to simulate the string
evolution in fig.\ \ref{fig:string}, we will take a short detour and
explain how to generate a string evolution as depicted in
fig.\ \ref{fig:boost}a.

First, let us notice that near the boundary, the metric is approximately AdS.
So the string evolves at very early times in a geometry which is AdS.
Moreover, in the rest frame of the string considered in
fig.\ \ref{fig:boost}a,
if the string has a sufficiently large energy, the string excitations
probe a geometry which is the Penrose limit of an in-falling massless
excitation in AdS. In other words, these excitations probe flat space.
To show this explicitly, we start by expanding around the null geodesic
$\bar x^\mu = (z,{\bm 0})$.
Defining, as in equations (\ref{eq:uformula}--\ref{eq:v}),
$\Delta x^\mu = x^\mu -\bar x^\mu(z), v=-\Delta x^\zero, u=-\Rads^2/z$,
yields
\begin{equation}
  ds^2
  = 2 \> du \> dv
    -\frac{dv^2}{z^2}
    + \sum_{i=\one,\two,\three}\frac{(d\Delta x^i)^2}{z^2}
  \,.
\end{equation}
Taking the Penrose limit by scaling the coordinates
as in (\ref{eq:scaling}) brings the metric into the Rosen form
\begin{equation}
  ds^2
  = 2 \> du \> dv
    + \sum_{i=\one,\two,\three}\frac{(d\Delta x^i)^2}{z^2}
  \,,
\end{equation}
which is just (\ref{eq:metricpp0}) with $\zh \to \infty$ (i.e.\ $f \to 1$).
While this is not immediately recognizable as a flat space metric,
a change of coordinates to bring the metric into Brinkmann form,
$\hat x^i=-u\,\Delta x^i/\Rads, \hat v=v-\sum_i (\Delta x^i)^2/(2u)$,
yields 
\begin{equation}
  ds^2= 2 \> du \> d\hat v + \sum_{i=\one,\two,\three}(d\hat x^i)^2 ,
\end{equation}
which is the $\zh \to \infty$ limit of (\ref{eq:ppmetrichat}).
In this geometry we consider the evolution of an initial string configuration
with the string shrunk to a point and with
$\partial_\tau \MT(0,\sigma) = A$,
$\partial_\tau\hat X^\three(0,\sigma) = B\cos\sigma$,
$\partial_\tau \MZ(0,\sigma) =
      \sqrt{(\partial_\tau \MT)^2-(\partial_\tau \hat X^\three)^2}$,
where we have defined $\MT$ and $\MZ$ by
$u=(\MT+\MZ)/2$ and $\hat v=\MT-\MZ$.
The other string coordinates remain zero throughout this section.
Since the string is point-like at the initial time, the stress-tensor
constraint $T_{\tau\sigma}=0$ is trivially satisfied, and the second stress
tensor constraint $T_{\tau\tau}=0$ was used to solve for the initial condition
on $\partial_\tau\MZ$. If we choose $A$ and $B$ such that  $B \ll A$,
the string has an approximate analytic solution of the form
$\MT=A\tau$,
$\hat X^\three=B\cos\sigma\sin\tau$,
$\MZ \approx A\tau$,
or, in the Brinkmann coordinates,
$U\approx A\tau$,
$\hat V\approx 0$,
$\hat X^\three=B\cos\sigma\sin\tau$.

\begin {figure}
\begin {center}
  \includegraphics[scale=0.5]{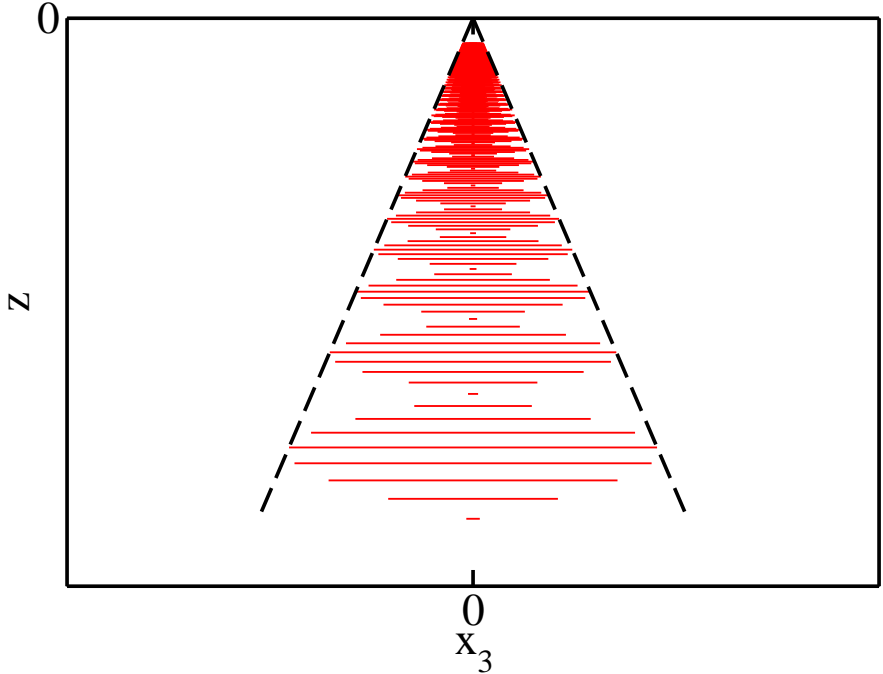}
  \caption{
     \label{fig:oscillate}
     An (approximate) AdS string solution for small $z$ showing the type
     of string we create with our initial conditions, viewed in the
     4-dimensional string rest frame.  The snapshots of the string
     are equally spaced in $\tau$.  $z_{\rm c}$ could be any point of
     the oscillation where the length of the string vanishes.  The
     straight line geodesics of fig.\ \ref{fig:boost}a may be roughly
     thought of
     as corresponding to the dashed lines above, which
     depict the envelope of the oscillation.
  }
\end {center}
\end {figure}

Next we switch to Rosen coordinates,
and we require that the string is again point-like at the initial time
$\tau=\tau_c$:
$\Delta X^\zero(\tau_c,\sigma)=0$,
$\Delta X^\three(\tau_c,\sigma)=0$ and
$Z(\tau_c,\sigma)=z_c$,
where we introduced a boundary regulator $z_c\approx 0$.
In order to satisfy this set of initial conditions we only have to
perform a shift in $U$ in the previous Brinkmann solution,
and use the mapping from Brinkmann to Rosen coordinates:
$U=-\Rads^2/Z \approx A(\tau-\tau_c)-\Rads^2/z_c$,
$V\approx 0$,
$X^\three = B(Z/\Rads)\cos\sigma\sin(\tau-\tau_c)$.
This is a solution which oscillates, as depicted in
fig.\ \ref{fig:oscillate} for small $z$.  The envelope of
these oscillations grows linearly with $z$ and roughly corresponds
to fig.\ \ref{fig:boost}a.

The next step is to perform a very large boost along $x^\three$ starting
from the rest frame we have been discussing so far. This will give us the
appropriate initial conditions
used to generate fig.\ \ref{fig:string}a.
Again, the string starts as a point-like object very close to the boundary:
$Z(\tau=0,\sigma)=z_c=0.01$,
$X^\zero(\tau,\sigma)=0$,
$X^\three(\tau,\sigma)=0$.
We take $z_h=1$.
The initial (worldsheet) time $\tau=0$ derivatives have the
following expressions:
$\partial_\tau X^\zero(0,\sigma)=1300+0.02\times 1299\cos\sigma$, 
$\partial_\tau X^\three(0,\sigma) =1299+0.02\times 1300\cos\sigma$,
$\partial_\tau Z(0,\sigma)
  =\sqrt{(\partial_\tau X^\zero)^2 f^2(z_c)- (\partial_\tau X^\three)^2 f(z_c)}$,
where $f(z_c)=1-(z_c/z_h)^4$.
Here $1300$ is the energy of the boosted string and $1299$ is the momentum
along the $x^\three$ direction in the appropriate units.

Earlier in this appendix, we discussed Penrose limits of AdS in order
to motivate our choice of initial conditions.
However, we now simulate the evolution of our classical string from
these initial conditions in the full AdS$_5$-Schwarzschild geometry,
with no such approximation, to obtain fig.\ \ref{fig:string}a.
In order to generate a string evolution to sufficiently late times,
we followed \cite{CheslerQuark, Herzog} and introduced an appropriate stretching
function $\stretch$,
\begin{equation}
  \stretch =
  \bigg(1+\frac{(X^\three)^2}{ z_h^2}\bigg)^m
  \bigg(\frac{z_h-z}{z_h-z_c}\bigg)^n
  \bigg(\frac{z_c}{z}\bigg)^p ,
\label{eq:stretching}
\end{equation}
with $m=-0.07$, $n=1$, $p=4$.
Then redefine the worldsheet time $\tau$ so that the worldsheet
metric is
\begin{equation}
  \gamma_{\alpha\beta}=\begin{pmatrix} -\stretch&0\\0&\stretch^{-1}
\end{pmatrix}.
\end{equation}
The numerical evolution of the classical
string was carried out using Mathematica.

Lastly, to generate fig.\ \ref{fig:string}b, the initial conditions were
modified as follows. The string begins its evolution once more as a
point-like object close to the boundary
$Z(\tau=0,\sigma)=z_c=0.01$, $X^\zero(\tau,\sigma)=0, X^\three(\tau,\sigma)=0$.
But the initial time $\tau=0$ derivatives are now
$\partial_\tau X^\zero(0,\sigma)=1300+0.6\times 1299 \cos\sigma, 
\partial_\tau X^\three(0,\sigma)=1299+0.6\times 1300 \cos\sigma$,
and, as before, the remaining
derivative is obtained from the constraint $T_{\tau\tau}=0$.

The main difference between the two initial conditions is that
we have increased the small parameter,
which is a placeholder for $\sqrt{(\alpha'/\Rads^2) \ln n_{\star}}$ from
0.02 in the previous numerical simulation to 0.6 here.
[As a side comment, this parameter cannot be increased to arbitrarily
large values or else the argument of the square root in $\partial_\tau Z$
becomes negative.]
The stretching function used here is of the same form as in 
(\ref{eq:stretching}), but with different exponents $m=-0.08, n=1, p=4.5$.
This string quickly stretches in $z$. Once one of the folding points reaches
the horizon, it remains frozen at that particular value of $x^\three$.
The net result is that we have now generated a string which is folded back,
with one folding point at the horizon and trailing behind the rest of the
string, while the other folding point is still close to the boundary.
This string is moving with some momentum in the $x^\three$ direction.


\end {document}